\documentclass[prl,twocolumn,superscriptaddress,longbibliography,nofootinbib]{revtex4-2}

\usepackage{amsmath,amssymb,amsfonts,bm}
\usepackage[colorlinks,citecolor=blue,linkcolor=red,urlcolor=blue]{hyperref}
\usepackage{xcolor}
\usepackage[normalem]{ulem}
\usepackage[dvipsnames]{xcolor}{\huge}
\usepackage{graphicx}
\usepackage{comment}

\DeclareMathOperator{\tr}{Tr}

\newcommand{\avg}[1]{\left\langle #1\right\rangle}
\newcommand{\bphi}{\bm{\phi}}
\newcommand{\btheta}{\bm{\theta}}
\newcommand{\sigmap}{{\sigma^\prime}}
\newcommand{\Vpp}{\ensuremath{V_{++}}}
\newcommand{\Vpm}{\ensuremath{V_{+-}}}
\newcommand{\OO}{\ensuremath{\mathcal{O}}}
\newcommand{\ket}[1]{\lvert #1\rangle}

\begin{document}

\title{Theory of criticality-enabled $U(1)$ symmetry breaking in a class of 1+1D systems}

\author{E. S. Andriyakhina}
\affiliation{Dahlem Center for Complex Quantum Systems, Fachbereich Physik, Freie Universit\"at Berlin, Arnimallee 14, 14195 Berlin}
\affiliation{Department of Physics, Massachusetts Institute of Technology, Cambridge, Massachusetts 02139, USA}
\author{A. S. Shankar}
\affiliation{The Abdus Salam International Center for Theoretical Physics (ICTP), Strada Costiera 11, I-34151 Trieste, Italy}
\author{T. Senthil}
\affiliation{Department of Physics, Massachusetts Institute of Technology, Cambridge, Massachusetts 02139, USA}
\author{Z. D. Shi}
\affiliation{Leinweber Institute for Theoretical Physics, Stanford University, Stanford, California 94305, USA}

\date{\today}

\begin{abstract}
We develop an analytic theory of the recently proposed $U(1)$ symmetry breaking at interacting 1+1D Lifshitz quantum critical points. The low-energy theory contains two compact fields, $\theta$ and $\phi$, coupled by a Berry term that makes $\theta$ canonically conjugate to the conserved density $\partial_x\phi$. Within a controlled large-$N$ limit, we establish long-range order (LRO) of $e^{i\beta\theta}$ and stretched-exponential decay of the conjugate vertex, $\log\langle e^{i\beta\phi(x)}e^{-i\beta\phi(0)}\rangle\propto-|\beta|^{4/3}|x|^{2/3}$. Beyond large $N$, we argue that $U(1)$ LRO follows if the dynamical exponent satisfies $z\neq1$, consistent with existing calculations. Using finite-size and infinite-system DMRG, we study an itinerant-electron chain with $U(1)\rtimes\mathbb{Z}_2$ symmetry, whose spin sector realizes this theory at a ferromagnetic quantum critical point. We find spin-nematic LRO in $S_i^+S_{i+1}^+$ and stretched-exponential electron Green's functions, revealing distinctive non-Gaussian signatures of criticality-enabled order.
\end{abstract}

\maketitle

\emph{Introduction.---} The fate of continuous symmetry breaking in low dimensions is a basic and subtle question in quantum many-body physics.  Standard infrared arguments suggest that long-wavelength Goldstone fluctuations are especially destructive in one spatial dimension. In their most familiar form, the Hohenberg--Mermin--Wagner--Coleman results rule out spontaneous breaking of continuous symmetries in closely related two-dimensional classical and relativistic settings \cite{MerminWagner1966,Mermin1967,Hohenberg1967,Coleman1973,Halperin2019_HMW_extension}. The corresponding one-dimensional quantum paradigm is the Luttinger liquid (LL), where logarithmic fluctuations of a compact phase field lead to power-law correlations and quasi-long-range order \cite{Haldane1981,Voit1995,Giamarchi2004}.
The intuition based on the LL is so robust that true long-range order of a continuous order parameter associated with an Abelian symmetry\footnote{Examples of spontaneously broken non-Abelian symmetries in one dimension have been constructed in Ref.~\cite{Fava2024_Heisenberg_randomsign}.} in a one-dimensional ground state is often regarded as impossible unless one introduces long-range interactions or other non-generic ingredients that require fine-tuning of infinitely many parameters in the Hamiltonian~\cite{Maghrebi2017,Ren2022, Ogunnaike2023_hydro,Fava2024_Heisenberg_randomsign,watanabe2024critical,Menon2024_MotzkinU1SSB,Sengoku2026_frustrationfreeSSB}.

There is, however, an important loophole. The usual intuition for the absence of continuous LRO in 1+1D quantum systems relies on a quantum-classical mapping that relates the 1+1D quantum system to a two-dimensional classical statistical model.\footnote{This intuition does not exclude orientational LRO in polymerized membranes, where nonlinear elasticity stabilizes a flat phase that breaks embedding-space rotations~\cite{LeDoussalRadzihovsky1992,Radzihovsky2003Membranes}.} Ferromagnets and related bosonic systems can instead possess a Berry-phase, or symplectic, term that is first order in imaginary time and makes the Euclidean weight complex in the density-phase variables. At long wavelengths, the Berry term makes the conserved $U(1)$ density canonically conjugate to the compact phase $\theta$ which shifts under $U(1)$ ($\theta$ is the azimuthal angle for spins or superfluid phase for bosons). This canonical structure modifies the infrared fluctuations of the would-be Goldstone mode.

\begin{figure}[t!]
    \centering\includegraphics[width=\columnwidth]{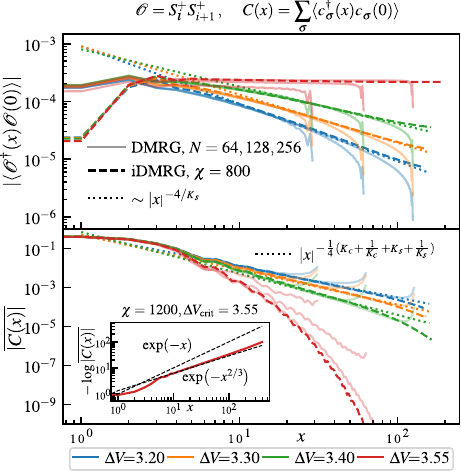}
    \caption{Numerically obtained correlation functions for the lattice model at the QCP (red, $\Delta V_{\mathrm{crit}} = 3.55$) and in the LL phase (blue, orange, and green, $\Delta V < \Delta V_{\mathrm{crit}}$). Dashed curves show iDMRG results, while faded solid curves show finite-size DMRG data for $N = 64, 128, 256$, with $x=0$ at the chain midpoint. In the LL phase, the correlators exhibit the expected power-law decay. \textbf{Top:} The nematic order parameter saturates at long distances, demonstrating true LRO at the QCP. \textbf{Bottom:} The electron Green's function $C(x)$, averaged over a four-site window. The inset provides a guide to the stretched-exponential decay, to be compared with the large-$N$ field-theory prediction $\log \chi_{\phi,\pi} \sim -x^{2/3}$.    }
    \label{fig:ShowLRO}
\end{figure}

The Berry term alone, however, does not guarantee LRO. The additional crucial ingredient is the \textit{critical softening} of the conjugate density: when its inverse uniform susceptibility vanishes, its leading quadratic cost becomes gradient-like. The long-wavelength zero-point fluctuations of $\theta$ can then become infrared finite, permitting true LRO at the quantum critical point (QCP).

A minimal field-theoretic realization of this idea was proposed by A. Nahum in Ref.~\cite{Nahum2025}. By bosonizing a spin-1 chain with the same symmetry ($U(1) \rtimes \mathbb{Z}_2$) as the XXZ model, Nahum derived a low energy Lifshitz field theory in the vicinity of a $\mathbb{Z}_2$ symmetry-breaking transition
\begin{equation}
\begin{aligned}
    \mathcal{L} = & i \partial_x \theta \, \partial_{\tau} \phi + \frac{1}{2} (\partial_x \theta)^2 + \frac{\lambda_2}{2} (\partial_x \phi)^2  \\
    & \quad + \frac{1}{2} (\partial_x^2 \phi)^2 + \frac{\lambda_4}{4} (\partial_x \phi)^4 + \dots\,, \label{eq:LE}
\end{aligned} 
\end{equation}
where $\theta$ and $\phi$ are compact $2\pi$- and 1-periodic dual fields, respectively. The Gaussian limit of Eq.~\eqref{eq:LE} also describes the transverse quantum fluid of an ideal superfluid edge dislocation~\cite{RadzihovskyEtAl2023TQF}, where continuous climb symmetry forbids local density interactions and permits $U(1)$ LRO. The new observation of Ref.~\cite{Nahum2025} is the emergence of internal $U(1)$ LRO precisely at a single-parameter-tuned, interacting QCP, where the quartic density interaction $(\partial_x\phi)^4$ is symmetry-allowed and relevant. Fixing $\lambda_4 > 0$ and tuning $\lambda_2$ from positive to negative drives the condensation of $S^z \sim \partial_x \phi$. The striking proposal of Ref.~\cite{Nahum2025}, based on controlled calculations of the scaling dimensions of $\phi$ and $\theta$, is that the $U(1)$ symmetry generated by $S^z$ is spontaneously broken precisely at the $\mathbb{Z}_2$ symmetry-breaking transition. However, since the Lifshitz field theory remains non-Gaussian at low energy, the physical $U(1)$ order parameter is a vertex operator $e^{i\beta\theta}$, whose functional form cannot be inferred directly from the scaling dimensions of $\phi$ and $\theta$. Therefore, a controlled analysis establishing $U(1)$ LRO has remained open.

In this Letter, we address the fate of $U(1)$ LRO and its physical consequences using a two-pronged approach. First, we study the Lifshitz critical point analytically. Assuming that the dynamical exponent $z$ satisfies $z \neq 1$ (strongly supported by existing numerical~\cite{Chepiga2026_numericsLifshitz} and analytical~\cite{Nahum2025,flores2025evidence} calculations), we show that the $U(1)$ symmetry must be spontaneously broken. To elucidate the structure of the critical state, we use a controlled large-$N$ version of the Lifshitz field theory to compute two-point correlators $\chi_{\theta,\beta}$/$\chi_{\phi,\beta}$ of both vertex operators $e^{i\beta\theta}$/$e^{i\beta\phi}$, where the allowed vertex charges $\beta$ are fixed by the compactification convention and chosen so that the vertices correspond to local operators in the microscopic realization. We show analytically that $\chi_{\theta,\beta}(x)$ approaches a nonzero constant at large separation, giving an explicit demonstration of $U(1)$ LRO at the critical point. By contrast, the conjugate vertex correlator $\chi_{\phi,\beta}(x)$ is neither long-ranged nor algebraic. After a short-distance Gaussian regime, its large-distance behavior is governed by a stretched exponential form $\log \chi_{\phi,\beta}(x)\sim - C_\phi |\beta|^{4/3} |x|^{2/3}$, up to a nonuniversal coefficient $C_\phi$. These signatures illustrate the strongly non-Gaussian behavior of the interacting Lifshitz theory.

Complementary to the analytic arguments, we also numerically study the itinerant $\mathbb{Z}_2$ ferromagnetic QCP in a lattice model
\begin{align}
    \!\!\!\!\!H \!=\! -t\!\sum_{i,\sigma}\! \left(c^\dagger_{i\sigma}c^{\phantom\dagger}_{i+1, \sigma} \!\!+\! \mathrm{h.c.}\right) \!\!+ \!U n_{i\uparrow}n_{i\downarrow} \!\!+ \!V_{\sigma\sigmap}n_{i\sigma}n_{i +\!1, \sigmap}.\!
    \label{eq:UV-ham}
\end{align}
As shown in Ref.~\cite{Yang2004}, the spin sector of the low-energy theory for this Hamiltonian maps exactly to the compact Lifshitz field theory with an operator correspondence 
\begin{equation}
    S^+_i S^+_{i+1} \sim e^{-2i\theta} \,, \quad c_{i, \uparrow} \sim \mathcal{O}_c \, e^{i(\pi\phi + \frac{\theta}{2})} \,,
\end{equation}
where $\mathcal{O}_c$ is an operator in the charge sector. Using (i)DMRG, we indeed find LRO in $S^+_i S^+_{i+1}$ and a stretched exponential decay in the correlation function of $c_{\uparrow}$, as shown in Fig.~\ref{fig:ShowLRO}. These findings confirm the qualitative validity of the large-$N$ theory even at $N = 1$. 

In what follows, we will begin with an analysis of the field theory and then present DMRG results on the itinerant fermion chain. We conclude by contrasting our results with other mechanisms for 1+1D $U(1)$ LRO and discuss interesting future directions.

\emph{Field theory.---}
We first focus on a generic field-theory description of systems governed by an interacting Lifshitz-like theory at the QCP. This description applies to various microscopic realizations, including one-dimensional itinerant electrons~\cite{SachdevSenthil1996,Yang2004,Kozii2017}, a spin-1 chain~\cite{Nahum2025}, and Lifshitz tricritical points in spin-1/2 chains, recently explored in Refs.~\cite{Rey2025,Rey2026}.
While our results are not restricted to any particular realization, for the specific case of itinerant electrons we discuss the physical implications of LRO in Sec.~``\textit{Lattice model}''.

We begin by deriving a nonperturbative constraint from anisotropic scale invariance. Starting from Eq.~\eqref{eq:LE}, integrating out $\theta$ generates a kinetic term $K (\partial_\tau \phi)^2$. Since interactions only depend on gradients of $\phi$, Wilsonian renormalization generates corrections to the effective Lagrangian that vanish at $k = 0$~\cite{SachdevSenthil1996,Nahum2025}. In particular, the coefficient $K$ of the kinetic term remains invariant. Assuming a general dynamical exponent $z$ at the QCP, the invariance of $K$ requires $\partial_x\phi$ to have scaling dimension $(3-z)/2$. On the other hand, if the $U(1)$ symmetry remains unbroken, the conserved $U(1)$ charge density $\partial_x\phi$ must have scaling dimension $1$~\cite{fisher1990presence,Wen1992_conserved_anisotropic,sachdev1994quantum}. Within a small-$\epsilon=2-d$ expansion of the Lifshitz field theory, Ref.~\cite{flores2025evidence} found $z=2-2\epsilon^2/243$, which extrapolates to $z\approx1.99$ at the physical value $\epsilon=1$, consistent with the numerical results of Ref.~\cite{Chepiga2026_numericsLifshitz}. Since $z\neq1$, the two scaling constraints on $\partial_x\phi$ are incompatible. Within these assumptions, the $U(1)$ symmetry must be spontaneously broken.

Although the argument above provides strong support for $U(1)$ LRO, it does not allow for controlled calculations of physical observables at the interacting QCP. To access these observables, we consider a large-$N$ generalization of the Lifshitz field theory in which $\theta$ and $\phi$ are promoted to $N$-component bosonic field vectors, $\btheta=(\theta_1,\ldots,\theta_N)$ and $\bphi=(\phi_1,\ldots,\phi_N)$, with Euclidean Lagrangian
\begin{equation}
    \mathcal L_0 = i\,\partial_x\btheta\cdot\partial_\tau\bphi + \frac{1}{2}(\partial_x\btheta)^2 + \frac{1}{2}(\partial_x^2\bphi)^2 + \frac{\lambda_2}{2}(\partial_x\bphi)^2 .
\label{eq:L0}
\end{equation}
The Berry-phase term is first order in imaginary time and makes $\partial_x\theta_a$ conjugate to $\phi_a$ (equivalently, $\theta_a$ is conjugate to $\partial_x\phi_a$). The most relevant interaction preserving the flavor symmetry and the shift symmetries of the fields is\footnote{For $N \neq 1$, the maximal flavor symmetry that preserves the compactification conditions is $O(N,\mathbb{Z}) \equiv (\mathbb{Z}_2^N) \rtimes S_N$. This symmetry allows for an additional relevant quartic term $\tilde \lambda_4 \sum_{a=1}^N (\partial_x \phi_a)^4$. The field theory with $\tilde \lambda_4 = 0$ therefore describes a multicritical point for $N \neq 1$.}
\begin{equation}
    \mathcal L_{\rm int} = \frac{\lambda_4}{4N}\left[(\partial_x\bphi)^2\right]^2,
\qquad \lambda_4>0 .
\label{eq:Lint}
\end{equation}
Eqs. \eqref{eq:L0} and \eqref{eq:Lint} are the large-$N$ analog of the Lifshitz action used for one-dimensional ferromagnetic quantum criticality in Eq.~\eqref{eq:LE}.  Integrating out $\btheta$ gives the Gaussian action
\begin{equation}
    S_0[\bphi]=\int d\tau d x\left[\frac{1}{2}(\partial_\tau\bphi)^2+\frac{1}{2}(\partial_x^2\bphi)^2+\frac{\lambda_2}{2}(\partial_x\bphi)^2 \right].
\label{eq:phi-action}
\end{equation}
The transition is tuned by the renormalized coefficient of $(\partial_x\bphi)^2$.  In a magnetic language, $\partial_x\phi$ is the density or longitudinal magnetization, while the appropriate local vertex $e^{i\beta\theta}$ is a transverse order parameter.

It is useful to decouple the interacting quartic term with a Hubbard-Stratonovich field $\psi$,
\begin{equation}
S[\bphi,\psi]=\frac{1}{2} \sum_{a=1}^N \int \phi_a g_\psi^{-1}\phi_a+\frac{N}{\lambda_4}\int \psi^2,
\label{eq:HS-action}
\end{equation}
where
\begin{equation}
    g_\psi^{-1}=-\partial_\tau^2+\partial_x^4-\lambda_2\partial_x^2-2i\partial_x(\psi\partial_x).
\label{eq:gpsi}
\end{equation}
After integrating out $\bphi$,
\begin{equation}
    S_{\rm eff}[\psi]=\frac{N}{\lambda_4}\int\psi^2+\frac{N}{2}\tr\log g_\psi^{-1} .
\label{eq:Seff}
\end{equation}
In the limit $N \to \infty$, the uniform saddle shifts the bare critical point. With a momentum cutoff $\Lambda$, the critical condition is
\begin{equation}
    \lambda_2^*+2i\psi_0=0, \qquad \lambda_2^*=-\lambda_4\frac{\Lambda}{2\pi},
\label{eq:critical-shift}
\end{equation}
so that the critical propagator is $g_0(\omega,k)=(\omega^2+k^4)^{-1}$. This results agrees with the large-$N$ calculations performed in~\cite{Nahum2025}.

\emph{Neutral vertex probes.---}
The compact vertex operators require a neutral insertion. We therefore study smeared correlators
\begin{equation}
    \chi_X[f] = \left\langle \exp\left[i\sum_{a=1}^N\int d\tau' d x' f(\tau',x')X_a(\tau',x')\right]\right\rangle,
\label{eq:chi-def}
\end{equation}
with $X=\phi$ or $\theta$ and the charge-neutrality condition $\int f=0$. The charge-$\beta$ two-point function $\chi_{X,\beta}(x)=\langle e^{i\beta\sum_a X_a(0,x)}e^{-i\beta\sum_a X_a(0,0)}\rangle$ is obtained from the sharp neutral source
\begin{equation} \label{eq:f_sharp}
    f(\tau', x') = \beta \delta(\tau')[\delta(x'-x) - \delta(x')].
\end{equation}
In the argument of the exponent in Eq.~\eqref{eq:chi-def}, we couple the probe equally to all $N$ components. For the sharp charge-$\beta$ insertion the flavor-space charge vector is $\bm{q}=\beta(1,\ldots,1)$, so that $\bm{q}^{\,2}=N\beta^2$. For fixed allowed $\beta=O(1)$, the source action is extensive in $N$ and modifies the leading large-$N$ saddle. This defines a heavy-vertex continuation that reduces to the physical neutral vertex at $N=1$.

To compute $\chi_{\theta, \beta}$, we write $f=\partial_x v$ and integrate out both $\btheta$ and $\bphi$. The source produces an additional term to the effective action in Eq.~\eqref{eq:Seff}, given by
\begin{equation}
    S_{\theta}[\psi]=\frac{N}{2}\left( \int v^2 - \int v(1)\,v(2)\,\partial_{\tau_1}\partial_{\tau_2}g_\psi(1,2)\right).
\label{eq:Sftheta}
\end{equation}
At the critical point, for the sharp probe given in Eq.~\eqref{eq:f_sharp}, we find that~\cite{SM}
\begin{equation}
    S_{\theta}[\psi_0] = N\beta^2\frac{\Lambda}{2\pi} \left(1-\frac{\sin \Lambda x}{\Lambda x}\right).
\label{eq:theta-constant}
\end{equation}
The source-induced correction is localized near the endpoints of the probe and its long-distance contribution decays as $|x|^{-2}$.  Consequently
\begin{equation}
    \ln \chi_{\theta,\beta}(x) = A_{\theta,\beta}\left[1+O(|x|^{-1})\right], \qquad x\to\infty,
\label{eq:theta-lro}
\end{equation}
where $A_{\theta,\beta}$ is cutoff- and charge-dependent.  This is true long-range order in $e^{i\beta\theta}$ at the critical point.

The density vertex behaves in the opposite way. For $X=\phi$, integrating out the $\bphi$ fields in the presence of the source adds
\begin{equation}
    S_{\phi}[\psi]=\frac{N}{2}\int f\,g_\psi\,f
\label{eq:Sfphi}
\end{equation}
to Eq.~\eqref{eq:Seff}.  Expanding about the critical saddle gives the source-induced displacement $\psi=\psi_0+\eta$.  The quadratic kernel for $\eta$ is $A(Q)=\lambda_4^{-1}+\Pi(Q)$, with
\begin{equation}
    \Pi(\Omega,q)=\frac{1}{4\sqrt{2}}\operatorname{Re}\frac{1}{\sqrt{q^2/2+i\Omega}},
\label{eq:Pi-main}
\end{equation}
so the Hubbard-Stratonovich response is intrinsically nonlocal.

At asymptotically large $x$, the expansion about the uniform critical saddle $\psi_0$ fails: for fixed nonzero $\beta$, the bare source contribution~\eqref{eq:Sfphi} scales as $\beta^2|x|$, whereas the leading correction induced by the deformed saddle scales as $\beta^4x^2$; see Ref.~\cite{SM} for details. The large-distance behavior is instead governed by a nonlinear source-dependent saddle. Introducing the response field
\begin{equation}
    u(\tau',x')=\int d\tau'' dx'' g_\psi(\tau'x';\tau''x'')f(\tau'', x''),
\end{equation}
and retaining the leading singular terms, the source-dependent action on the saddle reduces to
\begin{equation}
    \frac{-\log\chi_{\phi,\beta}(x)}{N} = \underset{u}{\operatorname{stat}} \int \left[fu - \frac{1}{2}(\partial_{\tau'}u)^2 - \frac{2\pi^2}{3}(\partial_{x'}u)^6 \right],
\end{equation}
where ``stat'' denotes evaluation at the stationary field $u = u^*$. For a sharp neutral source~\eqref{eq:f_sharp}, the rescaling
\begin{gather}
    x'=|x|X', \quad \tau'=|\beta|^{-2/3}|x|^{5/3}T', \\
    u(\tau',x')=\operatorname{sgn}(\beta)|\beta|^{1/3}|x|^{2/3} U(T',X') \label{eq:crit_scaling}
\end{gather}
makes the source, temporal, and nonlinear-gradient contributions to the action scale as $|\beta|^{4/3}|x|^{2/3}$. It follows that
\begin{equation}
    \log\chi_{\phi, \beta}(x) \sim -N C_\phi |\beta|^{4/3} |x|^{2/3},
\label{eq:chi_phi}
\end{equation}
where $C_\phi>0$ is nonuniversal. A direct numerical solution of the full source-deformed saddle equations is consistent with this asymptotic form; further details are provided in Ref.~\cite{SM}. 

Eq.~\eqref{eq:crit_scaling} also implies that the nonlinear saddle governing $\chi_{\phi,\beta}$ exhibits the anisotropic scaling $\tau\sim |x|^{5/3}$, corresponding to an effective dynamical exponent $z_\phi=5/3$ for this vertex response, distinct from the Lifshitz exponent $z=2$ associated with the $U(1)$ density response function.

Thus, the same interacting QCP simultaneously supports true phase LRO for the allowed local vertices $e^{i\beta\theta}$ and strong fluctuations of the corresponding conjugate vertices $e^{i\beta\phi}$.

\emph{Lattice model.---}
In order to further test the predictions of the large-$N$ field theory, we numerically study the model from Eq.~\eqref{eq:UV-ham}
in the canonical ensemble at a fixed $1/4$-filling of each spin species, using both open boundary condition density matrix renormalization group (DMRG)~\cite{white1992density,schollwock2011density} on finite chains of lengths $N$ = 64, 128 and 256, as well as in the thermodynamic limit using its infinite extension (iDMRG)~\cite{mcculloch2008infinite, phien2012infinite, karrasch2012luttinger}.  Throughout this Letter, we use parameterization $V_{\uparrow\uparrow} = V_{\downarrow\downarrow} = \Vpp$ and $V_{\uparrow\downarrow} = V_{\downarrow\uparrow} = \Vpm$; and keep $U=0.1, t=1$ fixed. Fixing $V_{++} = 0.8$ and varying $\Delta V = V_{+-} - V_{++}$, the model~\eqref{eq:UV-ham} exhibits a second order phase transition at $\Delta V_{\mathrm{crit}} \approx 3.55$ between a paramagnetic Luttinger liquid and a ferromagnet with a finite magnetization along the $z$-axis \footnote{An itinerant ferromagnetic Lifshitz critical point that breaks $SU(2)$ symmetry can instead be found in the Hubbard model with next-to-nearest neighbor hopping, as in Ref.~\cite{daul1998ferromagnetic}.}. On the lattice, the different operators in the microscopic theory can be related to two emergent bosonic fields $\phi_c$ and $\phi_s$ describing the charge and spin sector respectively through the bosonization dictionary (see \cite{SM} for details). In the LL phase, interactions in the charge and spin sectors can be parametrized using Luttinger parameters, which in the perturbative regime take the form $K_{c/s} = \left(1 \pm \frac{Ua}{\pi v_F} + \frac{2a(V_{++}\pm V_{+-})}{\pi v_F}\right)^{-1/2}$, where $a$ is the lattice spacing, which we set to $1$. We identify $\phi=\phi_s/\sqrt{2\pi}$,  $\theta=\sqrt{2\pi}\,\theta_s$ and $\lambda_2 = (2\pi/K_s)^2$.

The field $\phi_c$ remains a Luttinger liquid for all values of $\Delta V$ considered presently. 
The right moving part of the electron annihilation operator at site $i$ can be written as 
\begin{align}
    c_{i\uparrow} \sim e^{i(k_F x + \sqrt{\frac{\pi}{2}}(\phi_c+\phi_s+\theta_c+\theta_s))} \,.
\end{align}
The single-electron Green's function $C(x) = \sum_\sigma\avg{c^\dagger_\sigma(x) c^{\phantom\dagger}_\sigma(0)}$ exhibits Friedel oscillations at a characteristic wave-vector of $k_F = \pi/4$. 
The overall decay rate of single electron excitations is governed by the shortest ranged vertex operator. In the LL phase, all decay channels are power-law, and will all contribute, with the overall envelope given by 
$|x|^{-\frac{1}{4}\left(K_c + \frac{1}{K_c} + K_s + \frac{1}{K_s}\right)}.$
At the QCP, the $\phi_s$ operator is the shortest ranged, and 
the conjugate-vertex factor $\chi_{\phi,\pi}$ of the local electron operator, Eq.~\eqref{eq:chi_phi}, characterizes the leading spin-sector contribution to the decay. 

Consistently, DMRG reveals a clear transition from power-law decay in the LL phase to stretched-exponential decay at the critical point, with an exponent close to the large-$N$ nonlinear saddle prediction of $2/3$. This behavior is shown in Fig.~\ref{fig:ShowLRO} for the rolling-averaged correlator $\overline{C(x)}$, obtained using a four-site window to suppress the $k_F$ oscillations (see Ref.~\cite{SM} for details). 

\begin{figure}[t!]
    \centering
    \includegraphics[width=0.9\columnwidth]{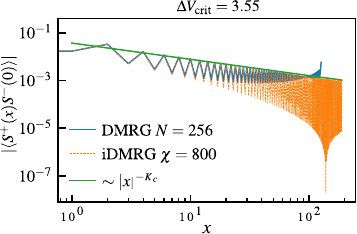}
    \caption{The correlation function of the operator $S^+$ at the QCP. The (i)DMRG numerics show characteristic $2k_F$ oscillations. This correlation function exhibits power-law decay due to the gapless charge mode. In a spin model without such a gapless charge sector, as in Refs.~\cite{Nahum2025, flores2025evidence}, the corresponding operator would instead exhibit true LRO.}
    \label{fig:SplusQLRO}
\end{figure}

The question of probing the LRO of a local $\theta$ vertex from Eq.~\eqref{eq:theta-lro} is more subtle on the lattice. Unlike in spin models, such as in Ref.~\cite{Nahum2025, flores2025evidence}, where $S^+$ maps directly to the appropriate local $\theta$ vertex (written as $e^{i\theta}$ in that convention), in the case of the itinerant electron system, described in Eq.~\eqref{eq:UV-ham}, the $S^+$ operator on the lattice picks up contributions from other vertex operators in the bosonized theory. Specifically, 
\begin{align}
    \!S^+_i\!\sim\!e^{-i\sqrt{2\pi}\theta_s}\left[\cos{\sqrt{2\pi}\phi_s}\!+\!\cos{(\sqrt{2\pi}\phi_c + 2k_Fx)}\right], 
\end{align}
meaning that even at the QCP, it has only quasi-long range order owing to the Luttinger-liquid character of the $\phi_c$ vertex, shown in Fig.~\ref{fig:SplusQLRO}. 
The simplest operator that exhibits true LRO should only depend on the $\theta_s$ vertex and is given by 
\begin{equation}
    \OO_i = S^+_i S^+_{i+1} \sim e^{-i\sqrt{8\pi}\theta_s}=e^{-2i\theta}.
\end{equation}
Thus this local order parameter corresponds to the charge $|\beta|=2$ phase vertex. This spin-2 operator breaks spin $U(1)$ symmetry down to $\mathbb{Z}_2$, and characterizes spin-nematic order. In the LL phase, the two-point function $\langle \OO_{i+x}^\dag \OO_{i} \rangle$ decays as $|x|^{-4/K_s}$. As one moves closer to the QCP, $K_s$ diverges, and one expects no decay at the critical point, implying true LRO. While this intuition is useful, it belies the non-Luttinger liquid nature of the critical point. Nonetheless, the long-range-ordered nature of $\OO$ is shown in Fig.~\ref{fig:ShowLRO}.

In summary, our numerical results suggest that, at the QCP, the itinerant-fermion system realizes an exotic ordered critical state with spin-nematic LRO. The state has no net longitudinal or transverse magnetization, $\langle S^z\rangle=\langle S^\pm\rangle=0$, and no dipolar transverse LRO: the conventional transverse-spin correlator remains quasi-long-range ordered, $\lim_{x\to\infty} \langle S_i^+S_{i+x}^-\rangle = 0$. By contrast, the spin-nematic correlator approaches a nonzero constant, $\lim_{x\to\infty} \langle \mathcal O_i^\dagger \mathcal O_{i+x}\rangle>0$, and exhibits true LRO.

\emph{Conclusion.---} In this Letter, we established through controlled analytic calculations and numerical simulations that an interacting quantum critical point (reached by tuning a single parameter) in one spatial dimension can support $U(1)$ LRO, thereby putting the proposal of Ref.~\cite{Nahum2025} on a rigorous footing. Given the simplicity of the itinerant fermion model that we numerically studied, it may be possible in the near future to experimentally realize this phenomenon in quantum simulator platforms.

Our result opens up many interesting future directions. One immediate target is to clarify the relation between the non-Gaussian Lifshitz field theory and the frustration-free $U(1)$ symmetry-breaking lattice models in Ref.~\cite{watanabe2024critical}. At first glance, since the frustration-free models have a Goldstone mode with $\omega \sim k^2$ dispersion, one may expect that they are equivalent to a multicritical point in the Lifshitz field theory, where both $\lambda_2$ and $\lambda_4$ are tuned to zero. However, this cannot be correct, as the multicritical Lifshitz theory has a unique ground state at finite system size $L$ accompanied by an Anderson tower of excited states, while the frustration-free models have an exact $O(L)$-ground state degeneracy. It is interesting to ask whether this is a fundamental distinction or a superficial one that can be eliminated by an infinitesimal perturbation of the frustration-free models.

A more ambitious goal is to identify the most general context in which criticality of one type of order can be used to enhance another type of order. Going beyond $U(1)$ symmetry, it would be fruitful to explore criticality-enabled non-Abelian continuous symmetry breaking in 1+1D, where Berry phase terms are replaced by more general Wess-Zumino-Witten terms. Since non-Abelian symmetry breaking is even more fragile by Mermin-Wagner arguments, this analysis should provide insights beyond the Abelian examples. 

\textit{Note added:} While this manuscript was in the final stages of preparation, Ref.~\cite{Chepiga2026_numericsLifshitz} appeared, reporting comprehensive numerical evidence for criticality-enabled $U(1)$ symmetry breaking in a spin-1 chain. Our work provides a complementary analytical treatment, including a controlled large-$N$ calculation, alongside numerical results for a distinct microscopic model involving itinerant fermions.

\begin{acknowledgments}
We thank Artem Kokovin, Pavel Nosov, Gal Lemut,  Sergey Khortsev, Michele Fabrizio, Hernan B. Xavier, and Zeno Bacciconi for useful discussions. The finite system DMRG calculations were performed using the iTensors.jl library~\cite{ITensor,ITensor-r0.3} and the iDMRG using TeNPy Library (version 1.1.0)~\cite{tenpy2024} on the \textit{Argo} cluster of the ICTP. The data and source code used to generate the figures in this letter are openly available at~\cite{zenodo_link}. AI tools (mainly Claude) were used as coding assistants in the numerical calculations. All generated code was carefully checked by the authors. ESA's work was supported by the Deutsche Forschungsgemeinschaft (DFG, German Research Foundation) through CRC-TR 183 (Project Number 277101999, subproject A02). TS was supported by the U.S. Department of Energy under Grant DE-SC0008739. ZDS was supported by a Leinweber Institute for Theoretical Physics postdoctoral fellowship at Stanford University and in part by the Gordon and Betty Moore Foundation EPiQS initiative, Grant GBMF8686.01.
\end{acknowledgments}

\bibliography{bib-Lifshitz}

\newpage
\phantom{}
\newpage
\begin{widetext}
\appendix

\begin{center}
{\large Supplemental Material for\\
``{Theory of criticality-enabled $U(1)$ symmetry breaking in a class of 1+1D systems}''}
\end{center}

\setcounter{figure}{0}
\renewcommand{\thefigure}{S\arabic{figure}}
\setcounter{page}{1}
\setcounter{section}{0}
\renewcommand{\thesection}{\Alph{section}}
\setcounter{equation}{0}
\renewcommand{\theequation}{S\arabic{equation}}
\setcounter{table}{0}
\renewcommand{\thetable}{S\arabic{table}}

Throughout the Supplemental Material we use the shorthand
\begin{equation}
    \int_P \equiv \int\frac{d\omega}{2\pi}\frac{dk}{2\pi},
    \qquad
    \int_Q \equiv \int\frac{d\Omega}{2\pi}\frac{dq}{2\pi},
\end{equation}
and we denote spacetime points by $y=(\tau,x)$.  Repeated flavor indices are summed.

\section{Effective field theory and large-$N$ saddle}
\label{app:effective-theory}

This appendix fixes conventions and derives the saddle-point equations used in the main text.  We start from Eqs.~\eqref{eq:L0} and \eqref{eq:Lint}, written as
\begin{equation}
\begin{aligned}
    S[\bm\theta,\bm\phi] = \int d\tau dx \bigg\{i\partial_x\bm\theta\cdot\partial_\tau\bm\phi +\frac12(\partial_x\bm\theta)^2 +\frac12(\partial_x^2\bm\phi)^2 +\frac{\lambda_2}{2}(\partial_x\bm\phi)^2 +\frac{\lambda_4}{4N}\left[(\partial_x\bm\phi)^2\right]^2\bigg\} .
\end{aligned}
\label{app:eq:theta-phi-action}
\end{equation}
The field $\partial_x\theta_a$ is canonically conjugate to $\phi_a$.  Completing the square and taking the Gaussian integral over $\bm\theta$ gives an overall determinant which is independent of $\bm\phi$.  Dropping this constant, we obtain the Lifshitz action
\begin{equation}
    S_{\rm b}[\bm\phi]=\int d\tau dx\left\{ \frac12(\partial_\tau\bm\phi)^2 + \frac12(\partial_x^2\bm\phi)^2 + \frac{\lambda_2}{2}(\partial_x\bm\phi)^2 + \frac{\lambda_4}{4N}\left[(\partial_x\bm\phi)^2\right]^2
\right\} .
\label{app:eq:phi-action}
\end{equation}

\noindent We decouple the quartic term by a Hubbard--Stratonovich field $\psi$,
\begin{equation}
    \exp\left[-\int d\tau dx\,\frac{\lambda_4}{4N}\left[(\partial_x\bm\phi)^2\right]^2\right] \propto \int {\cal D}\psi\, \exp\left[-\int d\tau dx\left(\frac{N}{\lambda_4}\psi^2+i\psi(\partial_x\bm\phi)^2\right)\right] .
\end{equation}
This gives
\begin{equation}
\begin{aligned}
    S_{\rm b}[\bm\phi,\psi] = \int d\tau dx\bigg\{\frac12(\partial_\tau\bm\phi)^2 +\frac12(\partial_x^2\bm\phi)^2 +\frac12(\lambda_2+2i\psi)(\partial_x\bm\phi)^2 +\frac{N}{\lambda_4}\psi^2 \bigg\}
\end{aligned}
\label{app:eq:HS-action}
\end{equation}
or, equivalently,
\begin{equation}
    S_{\rm b}[\bm\phi,\psi] =\frac12\iint dy_1dy_2\,\phi_a(y_1)\,g_\psi^{-1}(y_1,y_2)\,\phi_a(y_2) +\frac{N}{\lambda_4}\int dy\,\psi^2(y) .
\end{equation}
The inverse propagator is
\begin{equation}
    g_\psi^{-1}= -\partial_\tau^2+\partial_x^4-\lambda_2\partial_x^2-2i\partial_x(\psi\partial_x).
\label{app:eq:gpsi-inverse}
\end{equation}
With Fourier convention $\int\frac{d\omega}{2\pi}\frac{dk}{2\pi}\,e^{i\omega\tau+ikx}$, the kernel is
\begin{equation}
\begin{aligned}
    (g_\psi)_{\omega'k',\omega''k''}^{-1} =(2\pi)^2\delta(\omega'+\omega'')\delta(k'+k'')\left[\omega'^2+k'^4+\lambda_2k'^2\right] -2ik'k''\,\psi(\omega'+\omega'',k'+k'').
\end{aligned}
\label{app:eq:kernel-momentum}
\end{equation}
With this convention in mind, the action takes the following form,
\begin{equation}
    S_{\rm b}[\bm\phi,\psi] =\frac12\iint_{P_1, P_2} \phi_a(-P_1)\,(g_\psi)_{P_1,P_2}^{-1}\,\phi_a(-P_2) +\frac{N}{\lambda_4}\int_P \psi(P)\psi(-P) .
\end{equation}
Integrating out the $N$ components of $\bm\phi$ gives
\begin{equation}
    Z=\int{\cal D}\psi\,e^{-S_{\rm eff}[\psi]},
    \qquad
    S_{\rm eff}[\psi]=\frac{N}{\lambda_4}\int dy\,\psi^2(y)+\frac{N}{2}\tr\log g_\psi^{-1}.
\label{app:eq:Seff}
\end{equation}
The large-$N$ saddle is obtained from $\delta S_{\rm eff}/\delta\psi=0$.  Since
\begin{equation}
\frac{\delta g_\psi^{-1}(y_1,y_2)}{\delta\psi(y)}
=-2i\,\delta(\tau_1-\tau_2)\delta(x_1-x_2)\delta(\tau_1-\tau)\,
\partial_{x_1}\left[\delta(x_1-x)\partial_{x_1}\right],
\end{equation}
we find
\begin{equation}
    \frac{2}{\lambda_4}\psi(y)
    +i\,\partial_{x_1}\partial_{x_2}g_\psi(\tau,x_1;\tau,x_2)
    \big|_{x_1=x_2=x}=0.
\label{app:eq:saddle-general}
\end{equation}
For a uniform saddle $\psi(y)=\psi_0$ it is useful to define
\begin{equation}
    r_0\equiv\lambda_2+2i\psi_0,
    \qquad
    g_0(\omega,k)=\frac{1}{\omega^2+k^4+r_0k^2}.
\end{equation}
At the critical saddle $r_0=0$.  With a symmetric momentum cutoff $|k|<\Lambda$, Eq.~\eqref{app:eq:saddle-general} gives
\begin{equation}
    \frac{2}{\lambda_4}\psi_0+i\int_{-\Lambda}^{\Lambda}\frac{dk}{2\pi}\int\frac{d\omega}{2\pi}\frac{k^2}{\omega^2+k^4}=0,
\end{equation}
and therefore
\begin{equation}
    \psi_0=-i\lambda_4\frac{\Lambda}{4\pi},
    \qquad
    \lambda_2^*=-2i\psi_0=-\lambda_4\frac{\Lambda}{2\pi}.
\label{app:eq:critical-shift}
\end{equation}
This reproduces Eq.~\eqref{eq:critical-shift} of the main text.

\noindent The critical propagator is
\begin{equation}
    g_0(\tau,x; \tau', x')=\int_P\frac{e^{i\omega(\tau-\tau')+ik(x-x')}}{\omega^2+k^4}.
\label{app:eq:g0}
\end{equation}
At equal time,
\begin{equation}
\begin{aligned}
    g_0(0,0;0,0)-g_0(0,x;0,0) =\int_P\frac{1-e^{ikx}}{\omega^2+k^4} =\int\frac{dk}{2\pi}\frac{1-\cos(kx)}{2k^2} =\frac{|x|}{4}.
\end{aligned}
\label{app:eq:g0-difference}
\end{equation}
For the charge-$\beta$ two-point function, this gives the zeroth-order Gaussian result $\log\chi_{\phi,\beta}^{(0)}(x)=-N\beta^2|x|/4$ before the source-induced deformation of the saddle is included.

\section{The $\chi_{\phi,\beta}$ correlation function}
\label{app:chi-phi}

We now study the neutral vertex source functional
\begin{equation}
    \chi_\phi[f]
    =\left\langle \exp\left[i\sum_{a=1}^{N}\int dy\,f(y)\phi_a(y)\right]\right\rangle,
    \qquad
    \int dy\,f(y)=0 .
\label{app:eq:chi-phi-def}
\end{equation}
The generalized two-point function $\chi_{\phi,\beta}(x)=\langle e^{i\beta\sum_a\phi_a(0,x)}e^{-i\beta\sum_a\phi_a(0,0)}\rangle$ is obtained by choosing a neutral source whose two endpoint charges are $\pm\beta$, as in Eq.~\eqref{eq:f_sharp}. In the presence of the source the action is
\begin{equation}
\begin{aligned}
    S[\bm\phi]=-i\sum_a\int dy\,f(y)\phi_a(y) +\int dy\left\{\frac12(\partial_\tau\bm\phi)^2+\frac12(\partial_x^2\bm\phi)^2+\frac{\lambda_2}{2}(\partial_x\bm\phi)^2+\frac{\lambda_4}{4N}\left[(\partial_x\bm\phi)^2\right]^2\right\} .
\end{aligned}
\end{equation}
After the Hubbard--Stratonovich transformation and the Gaussian integral over $\bm\phi$, the effective action for $\psi$ becomes
\begin{equation}
    S[\psi]=S_{\rm eff}[\psi] + S_{\phi}[\psi],
    \qquad
    S_{\phi}[\psi]=\frac{N}{2}\iint dy_1dy_2\,f(y_1)g_\psi(y_1,y_2)f(y_2).
\label{app:eq:Sfphi}
\end{equation}
Using $\delta g=-g(\delta g^{-1})g$, the saddle equation in the presence of the source is
\begin{equation}
\begin{aligned}
    \frac{2}{\lambda_4}\psi(y) +i\,\partial_{x_1}\partial_{x_2}g_\psi(\tau,x_1;\tau,x_2)\big|_{x_1=x_2=x} -i\iint dy_1dy_2\,f(y_1)f(y_2) \left[\partial_x g_\psi(y_1;y)\right]\left[\partial_x g_\psi(y;y_2)\right] = 0.
\end{aligned}
\label{app:eq:source-saddle-phi}
\end{equation}
This is a nonlinear, nonlocal integral equation.  Below we first solve it perturbatively for a weak smooth source, and then explain why this perturbative solution fails at asymptotically large spatial separation.

\subsection{Perturbative solution of the source-deformed saddle}
\label{app:phi-saddle-pert}

Let
\begin{equation}
    \psi(y)=\psi_0+\eta(y),
    \qquad
    r_0=\lambda_2+2i\psi_0,
\end{equation}
where $\psi_0$ is the uniform bulk saddle.  The bulk effective action per flavor, $s_{\rm eff}\equiv S_{\rm eff}/N$, has the expansion
\begin{equation}
    s_{\rm eff}[\psi_0+\eta]
    =s_{\rm eff}[\psi_0]+\int_Q A(Q)\eta(Q)\eta(-Q)+\cdots,
\label{app:eq:bulk-expansion}
\end{equation}
with
\begin{equation}
    A(Q)=\lambda_4^{-1}+\Pi(Q).
\end{equation}
The polarization bubble is obtained from the second order term in the trace-log expansion.  Since
\begin{equation}
    \delta g_\psi^{-1}=-2i\partial_x(\eta\partial_x),
    \qquad
    \delta g_\psi^{-1}(P+Q,P)=2ik(k+q)\eta(Q),
\end{equation}
we have
\begin{equation}
\begin{aligned}
\Pi(\Omega,q)=\int_P
\frac{k^2(k+q)^2}
{\left(\omega^2+k^4+r_0k^2\right)
\left[(\omega+\Omega)^2+(k+q)^4+r_0(k+q)^2\right]}.
\end{aligned}
\label{app:eq:Pi-def}
\end{equation}
Performing the frequency integral gives
\begin{equation}
    \Pi(\Omega,q)=\int\frac{dk}{2\pi}
    \frac{k^2(k+q)^2}{2E_kE_{k+q}}
    \frac{E_k+E_{k+q}}{(E_k+E_{k+q})^2+\Omega^2},
    \qquad
    E_k=\sqrt{k^4+r_0k^2}.
\label{app:eq:Pi-after-omega}
\end{equation}
At criticality, $r_0=0$ and $E_k=k^2$.  Then
\begin{equation}
\begin{aligned}
    \Pi(\Omega,q)
    =\frac12\int\frac{dk}{2\pi}
    \frac{k^2+(k+q)^2}{\left[k^2+(k+q)^2\right]^2+\Omega^2} =\frac12\operatorname{Re}\int\frac{dp}{2\pi} \frac{1}{2p^2+q^2/2+i\Omega} =\frac{1}{4\sqrt2}\operatorname{Re}\frac{1}{\sqrt{q^2/2+i\Omega}} .
\end{aligned}
\label{app:eq:critical-bubble}
\end{equation}
In particular, the non-analyticity of $\Pi$, as a function of both $q$ and $\Omega$, is the simplest manifestation of the nonlocal response of the Hubbard--Stratonovich field.

The source term has the expansion
\begin{equation}
    s_{\phi}[\psi_0+\eta]
    =\frac12 fg_0f-i\int dy\,\eta(y)I_\phi(y)+\cdots,
\label{app:eq:source-expansion-phi}
\end{equation}
where $s_{\phi}=S_{\phi}/N$ and
\begin{equation}
    I_\phi(y) = i \frac{\delta s_{\phi}[\psi]}{\delta \psi}\bigg|_{\psi_0} = \iint dy_1dy_2\,f(y_1)\left[\partial_xg_0(y_1;y)\right]\left[\partial_xg_0(y;y_2)\right]f(y_2).
\label{app:eq:Iphi-real-space}
\end{equation}
Combining Eqs.~\eqref{app:eq:bulk-expansion} and \eqref{app:eq:source-expansion-phi}, the linearized saddle equation is
\begin{equation}
    2A(Q)\eta(Q)-iI_\phi(Q)=0, \qquad \eta(Q) = \frac{i}{2}\frac{I_\phi(Q)}{A(Q)}.
\label{app:eq:eta-solution}
\end{equation}
For example, in the static long-wavelength limit at the critical point,
\begin{equation}
    \eta(0, q)\simeq 2i|q| I_\phi(0, q),
    \qquad q\to0,
\end{equation}
The important point is the factor $|q|$: the saddle response is nonlocal and cannot be approximated by a local stiffness.

Substituting the saddle displacement back into the action gives
\begin{equation}
    \Delta s_{\phi,\rm fl}
    = \frac14\int_Q\frac{I_\phi(Q)I_\phi(-Q)}{\lambda_4^{-1}+\Pi(Q)}.
\label{app:eq:fluctuation-correction-phi}
\end{equation}
Thus the perturbative large-$N$ result is
\begin{equation}
    \chi_\phi[f]
    \simeq
    \exp\left[-N\left(\frac12 fg_0f
    +\frac14\int_Q\frac{I_\phi(Q)I_\phi(-Q)}{\lambda_4^{-1}+\Pi(Q)}
    \right)\right].
\label{app:eq:chi-phi-pert}
\end{equation}
For a concrete probe we take two smooth Gaussian packets separated by $x$ at equal time,
\begin{equation}
    f_{0x}^{(\beta)}(\tau',x')=\frac{\beta}{2\pi\sigma^2}
    \left[
    e^{-[(x'-x)^2+\tau'^2]/(2\sigma^2)}
    -e^{-[x'^2+\tau'^2]/(2\sigma^2)}
    \right].
\label{app:eq:gaussian-probe}
\end{equation}
From this point on, we assume $x>0$ without loss of generality. While our analytical results can be obtained in the limit $\sigma\to0$, reproducing Eq.~\eqref{eq:f_sharp} of the main text, our numerical simulations employ smooth neutral probes rather than sharp delta-function insertions. A delta-function source produces large local distortions of the saddle-point fields. In the present problem, the saddle-point equations are nonlinear and nonlocal, so such singular probes mix the desired long-distance response with short-distance large-deviation physics. As a result, the numerical problem becomes highly noisy and requires a much finer grid to achieve convergence. Smooth probes avoid this ultraviolet sensitivity while still resolving the separation between two well-defined insertions. A convenient choice is the difference between two smooth Gaussian packets localized around $(\tau,x)$ and $(0,0)$, as given in Eq.~\eqref{app:eq:gaussian-probe}. Its Fourier transform gives
\begin{equation}
\begin{aligned}
    \frac12fg_0f
    =2\beta^2\int\frac{dkd\omega}{(2\pi)^2}
    e^{-\sigma^2(k^2+\omega^2)}
    \frac{\sin^2(kx/2)}{\omega^2+k^4} \simeq
    \begin{cases}
    \beta^2 x/4, & x\gg\sigma,\\
    \beta^2 x^2/(16 \sqrt{\pi} \sigma), & \sigma\gg x,
    \end{cases}
\end{aligned}
\label{app:eq:gaussian-bare}
\end{equation}
The corresponding real-space source for the saddle is 
\begin{equation}
    I_\phi(\tau',x')
    = \beta^2\left[A(\tau', x'-x) - A(\tau', x') \right]^2,
\label{app:eq:Iphi-erf}
\end{equation}
where
\begin{equation}
    A(\tau', x') = \frac{1}{4} \int \frac{d \tau}{\sqrt{2\pi \sigma^2}} e^{-\frac{(\tau' - \tau)^2}{2 \sigma^2}} \mathrm{erf}\left(\frac{x'}{\sqrt{4 |\tau| + 2 \sigma^2}}\right) \,. 
\end{equation}
Importantly, the source term in Eq.~\eqref{app:eq:Sfphi} produces the bridge-like profile in Eq.~\eqref{app:eq:Iphi-erf}, which is extensive in the separation.  When evaluated on the undeformed critical saddle, the profile occupies a spatial region of length $L\sim x$ and an imaginary-time region of order $T\sim x^2$, as dictated by the bare $z=2$ propagator. This already shows that the source cannot be treated as a localized perturbation. The fully self-consistent nonlinear saddle has a different temporal scale, derived below.

Indeed, this perturbative approach is useful for smooth probes when $x$ is small, but it is not the correct asymptotic for the large-$x$ saddle. Rescaling $X' = x'/x$, $T' = \tau'/x^2$, and $\Sigma = \sigma/x \ll 1$, we would find that the Fourier transform of $I_\phi$ and the bubble propagator scale as
\begin{equation}
    I_\phi(\Omega, q) = \beta^2 x^3 \mathcal{I}_\phi(x^2 \Omega, xq), \qquad \Pi(\Omega,q)= x {\cal P}(x^2\Omega,x q).
\end{equation}
This implies
\begin{equation}
    \Delta s_{\rm fl} = \frac14\int_Q\frac{I_\phi(Q)I_\phi(-Q)}{\lambda_4^{-1}+\Pi(Q)} \approx \frac{\beta^4 x^2}{4} \int \frac{d\bar \Omega d \bar q}{(2\pi)^2} \frac{|\mathcal{I}_\phi(\bar\Omega, \bar q)|^2}{{\cal P}(\bar\Omega, \bar q)} \equiv \beta^4 x^2 C_{\phi,\mathrm{pert}},
\end{equation}
which is parametrically larger than $\frac12fg_0f \propto \beta^2 x$ in the same regime for fixed nonzero $\beta$, see Eq.~\eqref{app:eq:gaussian-bare}. The clean way to see the breakdown is to look at the induced saddle shift. Indeed, from Eq.~\eqref{app:eq:eta-solution}, we can estimate that
\begin{equation}
    r(Q) = 2i\eta(Q) = -\frac{I_\phi(Q)}{\lambda_4^{-1} + \Pi(Q)} \sim \beta^2 x^2,
\end{equation}
and hence in real space
\begin{equation}
    r(y) \sim \beta^2 x^{-1}.
\end{equation}
Although this correction vanishes pointwise as $x\to\infty$, it is not perturbatively small for the long-wavelength modes that dominate the observable. In Wigner representation and using the local approximation, if we insert this result into the adiabatic Green's function
\begin{equation}
    \bar g(y, y') = \int \frac{dk}{2\pi} \frac{d\omega}{2\pi} \frac{e^{i \omega(\tau - \tau') + ik(x - x')}}{\omega^2 + k^4 + r(\frac{y+y'}{2}) k^2}
\end{equation}
we would observe that in the critical propagator $\omega^2 + k^4 + r k^2$ the last two terms scale as
\begin{equation}
    k^4 \sim x^{-4}, \quad r k^2 \sim \beta^2 x^{-3},
\end{equation}
respectively, and hence $r k^2 \gg k^4$. Therefore the induced saddle shift is not perturbative for the long-wavelength modes that dominate the observable.

\begin{figure}[t!]
    \centering
    \includegraphics[width=0.99\textwidth]{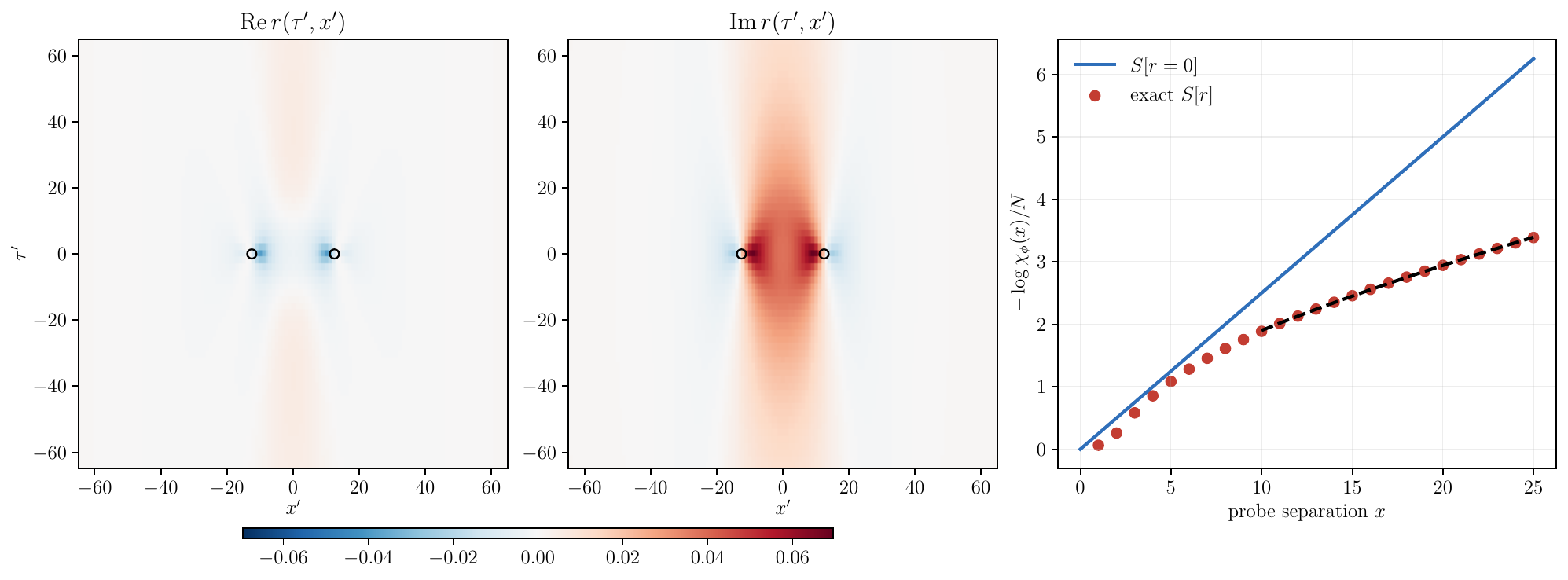}
    \caption{
    Representative source-induced saddle $r(\tau',x')=2i[\psi(\tau',x')-\psi_0]$ for a two-points probe with separation $x=25$; open circles mark the centers at $x'=\pm 12.5$, obtained from the exact numerical solution of Eq.~\eqref{app:eq:source-saddle-phi} on a periodic lattice. Left and middle: the real and imaginary parts reveal a strongly nonuniform complex saddle, including a pronounced profile connecting the two points. Right: the full numerical saddle (red) substantially lowers the action relative to the unmodified $r=0$ saddle (blue). The numerical source has $\beta=1$; the black dashed line is the large-$x$ fit $-\log\chi_{\phi,1}/N \propto x^{2/3}$, in agreement with the analytical prediction Eq.~\eqref{app:eq:phi-large-x-final}. The full discretized saddle equations were solved on a periodic $150\times150$ lattice with $L_{x'}=150$, $L_{\tau'}=300$, $\Delta x' = 1$, $\Delta \tau'=2$, $\lambda_4=1$. For our numerical simulation, we used the Gaussian packet given in Eq.~\eqref{app:eq:gaussian-probe} of the width $\sigma=1$. We solve the equation with five source-continuation steps and exact Newton iterations with relative tolerance $5\times10^{-4}$.}
    \label{fig:phi-nonlinear-saddle}
\end{figure}

\subsection{Large-distance nonlinear saddle}
\label{app:phi-large-x}

For the probe in Eq.~\eqref{app:eq:gaussian-probe}, the perturbative correction grows as $\Delta s_{\rm fl}\sim x^2$, whereas the bare Gaussian contribution grows only as $x$.  The expansion about the undeformed critical saddle is therefore nonuniform in $x$.  We now derive the self-consistent asymptotic saddle.

We write the source-induced stiffness as
\begin{equation}
    r(y)=2i[\psi(y)-\psi_0]
\label{app:eq:r-def}
\end{equation}
and define
\begin{equation}
    \mathcal L_r=-\partial_{\tau'}^2+\partial_{x'}^4-\partial_{x'}\!\left[r(y')\partial_{x'}\right],
    \qquad g_r=\mathcal L_r^{-1}.
\label{app:eq:Lr-def}
\end{equation}
For fixed $r$, introduce the classical response field
\begin{equation}
    u(y')=\int dy''\,g_r(y',y'')f(y''),
    \qquad \mathcal L_r u=f.
\label{app:eq:u-def}
\end{equation}
The nonlocal source term admits the exact stationary representation
\begin{equation}
\begin{aligned}
    \frac12 f g_r f
    =\underset{u}{\operatorname{stat}}\Bigg\{
    \int dy'\,f u
    -\frac12\int dy'\,
    \left[(\partial_{\tau'} u)^2+(\partial_{x'}^2u)^2+r(\partial_{x'}u)^2\right]
    \Bigg\}.
\end{aligned}
\label{app:eq:source-legendre}
\end{equation}
Here $\operatorname{stat}$ denotes evaluation at the stationary field $u=g_rf$.  Equation~\eqref{app:eq:source-legendre} follows directly by completing the square, or by varying the expression in braces to obtain Eq.~\eqref{app:eq:u-def}.

At large $x$, the saddle is slowly varying on the correlation scales generated by $r$, as verified below.  We may therefore use the leading local, or adiabatic, expansion of the trace-log.  For a uniform small stiffness, the singular part of the bulk effective action per flavor is
\begin{equation}
\begin{aligned}
    \mathcal F_{\rm b}[r]
    = -\frac{r^2}{4\lambda_4}
    +\frac12\int_P\left[
    \log\left(1+\frac{r k^2}{\omega^2+k^4}\right)
    -\frac{r k^2}{\omega^2+k^4}\right] = -\frac{r^{3/2}}{6\pi}+O(r^2).
\end{aligned}
\label{app:eq:bulk-free-energy-local}
\end{equation}
The subtraction after the first equality sign implements the critical-point condition in Eq.~\eqref{app:eq:critical-shift}: the linear ultraviolet contribution from the determinant is canceled by the Hubbard--Stratonovich term. Indeed, from $\psi(y) = \psi_0 - ir(y)/2$, we find that
\begin{align}
   s_{\rm eff}[r] - s_{\rm eff}[r=0] & = \frac{1}{4\lambda_4} \int dy [(\psi_0 - ir/2)^2 - \psi_0^2] + \frac{1}{2} \int dy \int_P [\log g_r^{-1} - \log g_{r=0}^{-1}] \notag \\
   & = - \int dy \frac{i r \psi_0}{\lambda_4} + \frac12 \int dy \int_P \frac{r k^2}{\omega^2+k^4}  + \int dy \, \mathcal F_{\rm b}[r(y)] \notag \\
   & = - \frac{\Lambda}{4 \pi} \int dy\, r + \frac{\Lambda}{4 \pi} \int dy\, r + \int dy \, \mathcal F_{\rm b}[r(y)] =  \int dy \, \mathcal F_{\rm b}[r(y)]
\end{align}
where we used the critical-point value $\psi_0 = -i \lambda_4 \Lambda/(4\pi)$ from Eq.~\eqref{app:eq:critical-shift}. The fractional power in Eq.~\eqref{app:eq:bulk-free-energy-local} is understood on the branch obtained by analytic continuation from the critical Hubbard--Stratonovich saddle.

Combining Eqs.~\eqref{app:eq:source-legendre} and \eqref{app:eq:bulk-free-energy-local}, the leading local functional for the source-dependent action per flavor is
\begin{equation}
\begin{aligned}
    \Delta s_\phi[u,r]=\int dy'\Bigg[&fu
    -\frac12(\partial_{\tau'} u)^2
    -\frac12(\partial_{x'}^2u)^2
    -\frac12r(\partial_{x'}u)^2 -\frac{r^{3/2}}{6\pi}
    +O(r^2,\partial r)\Bigg],
\end{aligned}
\label{app:eq:local-ur-action}
\end{equation}
with $\log\chi_{\phi,\beta}=-N\Delta s_{\phi}$ evaluated at the joint saddle for the charge-$\beta$ source.  Stationarity with respect to $r$ gives the local saddle condition
\begin{equation}
    (\partial_{x'}u)^2+\frac{\sqrt r}{2\pi}=0.
\label{app:eq:local-gap-equation}
\end{equation}
The root continuously connected to $r=0$ obeys
\begin{equation}
    \sqrt r=-2\pi(\partial_{x'}u)^2,
    \qquad
    r=4\pi^2(\partial_{x'}u)^4.
\label{app:eq:r-of-u}
\end{equation}
Substituting Eq.~\eqref{app:eq:r-of-u} back into Eq.~\eqref{app:eq:local-ur-action} gives
\begin{equation}
\begin{aligned}
    \Delta s_\phi
    =\underset{u}{\operatorname{stat}}\int d\tau' dx' \Bigg[
    fu-\frac12(\partial_{\tau'} u)^2
    -\frac12(\partial_{x'}^2u)^2 -\frac{2\pi^2}{3}(\partial_{x'}u)^6 \Bigg].
\end{aligned}
\label{app:eq:local-u-action}
\end{equation}
The corresponding nonlinear saddle equation is
\begin{equation}
    -\partial_{\tau'}^2u+\partial_{x'}^4u
    -4\pi^2\partial_{x'}\!\left[(\partial_{x'}u)^5\right]=f.
\label{app:eq:nonlinear-u-equation}
\end{equation}
The large-distance scaling is most transparent for the sharp neutral source
\begin{equation}
    f_{x,\beta}(\tau',x')=\beta\,\delta(\tau')\left[\delta(x'-x)-\delta(x')\right].
\label{app:eq:sharp-phi-source}
\end{equation}
A fixed-width smooth probe has the same asymptotic scaling because its width becomes negligible compared with $x$. For any fixed nonzero allowed $\beta$, the powers of $x$ are unchanged. Although this nonlinear differential equation cannot be solved exactly, we can obtain the asymptotic form of its solution at large $x$ and $\tau$ through a scaling analysis. Observe first that the LHS contains three terms with different numbers of spacetime derivatives and different powers of $u$. Under a general scaling Ansatz $x' = xX', \tau' = x^{\alpha_{\tau}} T', u(\tau', x') = x^{\alpha_u} U(T', X')$, it is impossible to arrange the exponents $\alpha_{\tau}, \alpha_u$ such that every term on the LHS carries the same power of $x$ as the source term on the RHS. At best, one can arrange so that two of the terms on the LHS have the same power of $x$ as the source, while the third term is subleading. To determine the optimal scaling, we turn to a casework:
\begin{enumerate}
    \item If terms 1 and 2 on the LHS balance the RHS, while term 3 on the LHS is subleading, we need to impose the constraints
    \begin{equation}
        - 2 \alpha_{\tau} + \alpha_u = -4 + \alpha_u = - \alpha_{\tau} - 1 > - 6 + 5 \alpha_u \,. 
    \end{equation}
    The chain of equalities admits a unique solution $\alpha_{\tau} = 2, \alpha_u = 1$. However, this solution does not satisfy the last inequality. Thus, this possibility cannot be self-consistent. 
    \item If terms 2 and 3 on the LHS balance the RHS, while term 1 on the LHS is subleading, we need to impose the constraints
    \begin{equation}
        -4 + \alpha_u = -6 + 5 \alpha_u = - \alpha_{\tau} - 1 > - 2 \alpha_{\tau} + \alpha_u \,,
    \end{equation}
    which admits a self-consistent solution $\alpha_{\tau} = 5/2, \alpha_u = 1/2$. The saddle point action scales as $\Delta s_{\phi}[u] \sim x^{1/2}$ in the large $x$ limit and higher order gradient terms are suppressed since $\alpha_u < 1$. 
    \item If terms 1 and 3 on the LHS balance the RHS, while term 2 on the LHS is subleading, we need to impose the constraints
    \begin{equation}
        -2\alpha_{\tau} + \alpha_u = - 6 + 5 \alpha_u = - \alpha_{\tau} - 1 > - 4 + \alpha_u \,,
    \end{equation}
    which admits a different self-consistent solution $\alpha_{\tau} = 5/3, \alpha_u = 2/3$. The saddle point action scales as $\Delta s_{\phi}[u] \sim x^{2/3}$ in the large $x$ limit and higher order gradient terms are suppressed since $\alpha_u < 1$. 
\end{enumerate}
From this casework, we see that there are two possible scaling hypotheses that can be self-consistent. Upon taking the large $x$ limit, the third Ansatz with $\Delta s_{\phi}[u] \sim x^{2/3}$ dominates over the second Ansatz with $\Delta s_{\phi}[u] \sim x^{1/2}$. Therefore, from now on, we will work with the third Ansatz. Under the change of variables
\begin{equation}
    x'=xX',
    \qquad
    \tau'=|\beta|^{-2/3}x^{5/3}T',
    \qquad
    u(\tau',x')=\operatorname{sgn}(\beta)|\beta|^{1/3}x^{2/3}U(T',X'),
\label{app:eq:large-L-rescaling}
\end{equation}
the temporal term, the nonlinear spatial term, and the source in Eq.~\eqref{app:eq:nonlinear-u-equation} all scale in magnitude as $|\beta|^{5/3}x^{-8/3}$.  The Lifshitz term $\partial_{x'}^4u$ scales in magnitude as $|\beta|^{1/3}x^{-10/3}$ and is thus subleading at fixed nonzero $\beta$. The leading dimensionless saddle therefore satisfies
\begin{equation}
    -\partial_{T'}^2U   -4\pi^2 \partial_{X'} \!\left[(\partial_{X'} U)^5\right]
    =\delta(T')\left[\delta(X'-1)-\delta(X')\right].
\label{app:eq:universal-nonlinear-saddle}
\end{equation}
The stiffness profile scales as
\begin{equation}
    r(\tau',x')=|\beta|^{4/3}x^{-4/3}R(T',X'),
    \qquad
    R(T', X') = 4\pi^2(\partial_{X'} U)^4+\cdots
\label{app:eq:r-profile-scaling}
\end{equation}
As briefly mentioned before, this scaling also verifies the adiabatic approximation.  A uniform stiffness $r$ generates the crossover scales
\begin{equation} \label{app:eq:xi_x_xi_tau}
    \xi_{x'}\sim |r|^{-1/2}\sim |\beta|^{-2/3}x^{2/3},
    \qquad
    \xi_{\tau'}\sim |r|^{-1}\sim |\beta|^{-4/3}x^{4/3}.
\end{equation}
The saddle varies over $x$ in space and $|\beta|^{-2/3}x^{5/3}$ in imaginary time, see Eq.~\eqref{app:eq:r-profile-scaling}, and hence
\begin{equation}
    \frac{\xi_{x'}}{x}\sim
    \frac{\xi_{\tau'}}{|\beta|^{-2/3}x^{5/3}}\sim |\beta|^{-2/3}x^{-1/3} \to 0.
\label{app:eq:adiabatic-control}
\end{equation}
Derivative corrections to the local determinant are therefore asymptotically suppressed. Likewise, the omitted $\int(\partial_{x'}^2u)^2$, $\int(\partial_{x'}u)^8$, and $\int r^2$ contributions are only $O(x^0)$, whereas the leading action is $O(x^{2/3})$.

It follows that
\begin{equation}
    \Delta s_{\phi,\beta}(x)=C_\phi |\beta|^{4/3}x^{2/3}+o(x^{2/3}),
\label{app:eq:phi-action-scaling}
\end{equation}
where the dimensionless coefficient is determined by the universal nonlinear capacity problem
\begin{equation}
\begin{aligned}
    C_\phi=\underset{U}{\operatorname{stat}}\Bigg\{
    U(0,1)-U(0,0)
    -\int dT'\,dX'\left[
    \frac12(\partial_{T'}U)^2
    +\frac{2\pi^2}{3}(\partial_{X'}U)^6
    \right]\Bigg\}.
\end{aligned}
\label{app:eq:Cphi-functional}
\end{equation}
Consequently, at large distances,
\begin{equation}
    \log\chi_{\phi,\beta}(x)
    =-N C_\phi |\beta|^{4/3}x^{2/3},
    \qquad C_\phi>0.
\label{app:eq:phi-large-x-final}
\end{equation}
In the normalization used here, $C_\phi$ is fixed by the dimensionless problem in Eq.~\eqref{app:eq:Cphi-functional}.  Restoring the coefficients of the continuum action and the microscopic normalization of the vertex makes the overall coefficient nonuniversal, while the exponent is fixed by the homogeneous infrared saddle.

A direct numerical solution of the source-deformed saddle equations is consistent with this prediction.  A representative numerical stiffness profile is shown in Fig.~\ref{fig:phi-nonlinear-saddle}; it is strongly inhomogeneous and forms an extended bridge between the two insertions. The corresponding $\beta=1$ saddle action, shown in the right panel of Fig.~\ref{fig:phi-nonlinear-saddle}, is well described at large separation by $-\log\chi_{\phi,1}/N\propto x^{2/3}$; Eq.~\eqref{app:eq:phi-large-x-final} gives the result for general allowed $\beta$.

\section{The $\chi_{\theta,\beta}$ correlation function}
\label{app:chi-theta}

\begin{figure}[t!]
    \centering
    \includegraphics[width=0.99\textwidth]{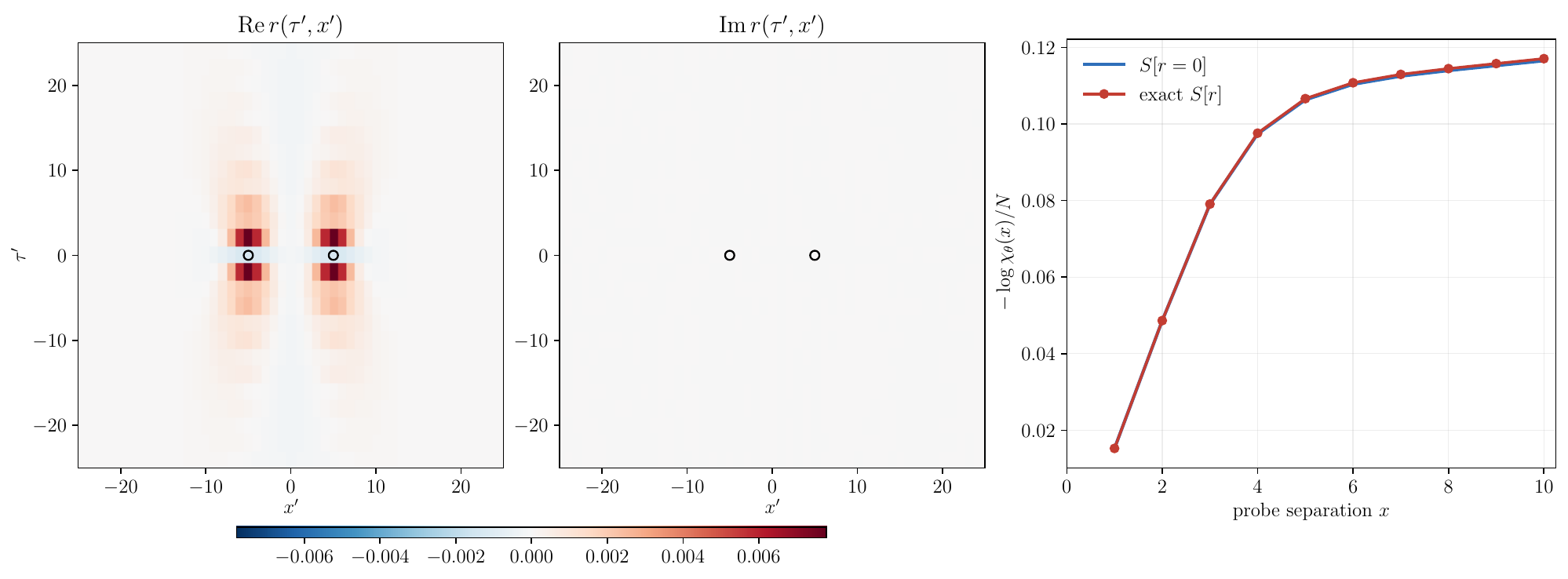}
    \caption{
    Representative source-induced saddle $r(\tau',x')=2i[\psi(\tau',x')-\psi_0]$ for a two-point probe with separation $x=10$. The open circles mark the probe centers at $x'=\pm 5$. The saddle was obtained from the exact numerical solution of Eq.~\eqref{app:eq:source-saddle-theta} on a periodic lattice. Left and middle: the saddle is purely real and deviates only slightly from the uniform bare solution, $r=0$. Right: comparison of the full numerical saddle (red) with the bare saddle (blue), illustrating that the two are nearly indistinguishable. The full discretized saddle-point equations were solved on a periodic $150\times150$ lattice with $L_{x'}=150$, $L_{\tau'}=300$, $\Delta x'=1$, $\Delta\tau'=2$, and $\lambda_4=1$. The Gaussian probe, Eq.~\eqref{app:eq:gaussian-probe}, had width $\sigma=1$ and charge $\beta=1$. The equations were solved using five source-continuation steps followed by exact Newton iterations with a relative tolerance of $10^{-7}$.}
    \label{fig:theta-nonlinear-saddle}
\end{figure}

We next compute the neutral source functional for the conjugate phase field,
\begin{equation}
    \chi_\theta[f]
    =\left\langle \exp\left[i\sum_{a=1}^{N}\int dy\,f(y)\theta_a(y)\right]\right\rangle,
    \qquad
    \int dy\,f(y)=0 .
\label{app:eq:chi-theta-def}
\end{equation}
The generalized two-point function $\chi_{\theta,\beta}(x)=\langle e^{i\beta\sum_a\theta_a(0,x)}e^{-i\beta\sum_a\theta_a(0,0)}\rangle$ is obtained by using the same endpoint charges $\pm\beta$ as in Eq.~\eqref{eq:f_sharp}. For a neutral source we may write
\begin{equation}
    f(\tau,x)=\partial_x v(\tau,x),
    \qquad
    v(\tau,x)=\int_{-\infty}^{x}dx'\,f(\tau,x'),
\label{app:eq:v-def}
\end{equation}
with $v\to0$ at $x\to\pm\infty$.  The source term becomes, after integrating by parts,
\begin{equation}
    -i\int dy\,f(y)\theta_a(y)=i\int dy\,v(y)\partial_x\theta_a(y).
\end{equation}
After integrating out $\bm\theta$, the source-deformed $\bm\phi$ action contains the linear term $-v\partial_\tau\phi_a$.  Repeating the Hubbard--Stratonovich decoupling and integrating over $\bm\phi$ gives
\begin{equation}
    S[\psi]=S_{\rm eff}[\psi]+S_{\theta}[\psi],
\end{equation}
where
\begin{equation}
    S_{\theta}[\psi]
    =\frac{N}{2} \int dy \, v(y)^2 -\frac{N}{2}\iint dy_1dy_2\,
    v(y_1)v(y_2)\,
    \partial_{\tau_1}\partial_{\tau_2}g_\psi(y_1,y_2).
\label{app:eq:Sftheta}
\end{equation}
This is Eq.~\eqref{eq:Sftheta} of the main text.  The corresponding saddle equation is
\begin{equation}
\begin{aligned}
    \frac{2}{\lambda_4}\psi(y) +i\,\partial_{x_1}\partial_{x_2}g_\psi(\tau,x_1;\tau,x_2)\big|_{x_1=x_2=x} +i\iint dy_1dy_2\,v(y_1)v(y_2) \left[\partial_{\tau_1}\partial_x g_\psi(y_1;y)\right] \left[\partial_{\tau_2}\partial_x g_\psi(y;y_2)\right] = 0.
\end{aligned}
\label{app:eq:source-saddle-theta}
\end{equation}

\subsection{Perturbative source response}
\label{app:theta-saddle-pert}

The source term in the linearized saddle equation is
\begin{equation}
    I_\theta(y) =  i \frac{\delta s_{\theta}[\psi]}{\delta \psi}\bigg|_{\psi_0} = - \iint dy_1dy_2\,v(y_1)v(y_2) \left[\partial_{\tau_1}\partial_xg_0(y_1;y)\right] \left[\partial_{\tau_2}\partial_xg_0(y;y_2)\right].
\label{app:eq:Itheta-real-space}
\end{equation}
For the same two-packet probe as in Eq.~\eqref{app:eq:gaussian-probe},
\begin{equation}
    v(\tau',x')=\beta\,\frac{e^{-\tau'^2/(2\sigma^2)}}{2\sqrt{2\pi}\sigma}
    \left[
    \operatorname{erf}\left(\frac{x'-x}{\sqrt{2\sigma^2}}\right)
    -\operatorname{erf}\left(\frac{x'}{\sqrt{2\sigma^2}}\right)
    \right].
\label{app:eq:v-gaussian}
\end{equation}
In the sharp-probe limit, $\sigma \to 0$, this becomes
\begin{equation}
    v(\tau',x') = -\beta\,\delta(\tau')\Theta(x')\Theta(x-x') , \quad v(\omega, k) = -\beta\,e^{-ikx/2} \frac{2\sin(k x/2)}{k}.
\label{app:eq:v-sharp}
\end{equation}
Using the critical propagator, one obtains
\begin{equation}
    I_\theta(\tau',x')
    = -\beta^2\left[
    \frac{e^{-(x'-x)^2/(4|\tau'|)}-e^{-x'^2/(4|\tau'|)}}{4\sqrt{\pi|\tau'|}}
    \right]^2 .
\label{app:eq:Itheta-endpoints}
\end{equation}
At the unperturbed critical saddle,
\begin{equation}
\begin{aligned}
    s_{\theta}[\psi_0] =\frac12\int_P |v(\omega, k)|^2 \left(1 - \frac{\omega^2}{\omega^2 + k^4} \right) = \beta^2\int_{-\Lambda}^{+\Lambda} \frac{dk}{2\pi} \sin^2\left(\frac{k x}{2}\right) = \beta^2\frac{\Lambda}{2\pi} \left(1-\frac{\sin \Lambda x}{\Lambda x}\right) \xrightarrow{x\Lambda\gg1} \beta^2\frac{\Lambda}{2\pi}.
\end{aligned}
\label{app:eq:sftheta-constant}
\end{equation}
The constant is ultraviolet dependent and contributes to the nonuniversal amplitude of the ordered correlator.  Its saturation with $x$ is the important point.

The fluctuation correction has the same structure as before,
\begin{equation}
    \Delta s_{\theta,\rm fl}
    =\frac14\int_Q\frac{I_\theta(Q)I_\theta(-Q)}{\lambda_4^{-1}+\Pi(Q)}.
\label{app:eq:theta-fl-correction}
\end{equation}
We can again find its scaling with $x$. Using
\begin{align}
    I_\theta (\tau', x') = \beta^2 x^{-2} \mathcal{I}_\theta (x^2 T', x X'), \qquad I_\theta(\Omega, q) = \beta^2 x \widetilde{\mathcal{I}}_\theta(x^2 \Omega, x q), \qquad \Pi(\Omega,q)= x {\cal P}(x^2\Omega,x q),
\end{align}
we find that
\begin{equation}
    \Delta s_{\rm fl} - \Delta s_{\rm fl}^{\rm UV} \approx \beta^4 x^{-2} C_\theta,
    \label{app:eq:theta-xminus2}
\end{equation}
where $\Delta s_{\rm fl}^{\rm UV} \approx 2.1 \times 10^{-3}\beta^4 \lambda_4 \Lambda$ is a small $x$-independent constant (the quoted coefficient is for the normalization used here).  Combining Eqs.~\eqref{app:eq:sftheta-constant} and \eqref{app:eq:theta-xminus2}, we obtain
\begin{equation}
    \ln \chi_{\theta,\beta}(x) = A_{\theta,\beta}\left[1+O(x^{-1})\right], \qquad A_{\theta,\beta}\neq0.
\label{app:eq:theta-lro-final}
\end{equation}
Thus, the allowed phase vertex $e^{i\beta\theta}$ exhibits true long-range order at the critical point, whereas the conjugate vertex $e^{i\beta\phi}$ is suppressed by the large-distance nonlinear saddle described above. We verify this result by comparison with the exact numerical solution of Eq.~\eqref{app:eq:source-saddle-theta}. Figure~\ref{fig:theta-nonlinear-saddle} shows a representative source-induced saddle and compares the action evaluated on the exact saddle with that evaluated on the bare saddle ($r=0$), demonstrating good agreement between the two.

\section{Numerical solution of the saddle equations}
\label{app:numerics}

\begin{figure}[t!]
\centering
    \includegraphics[width=0.45\linewidth]{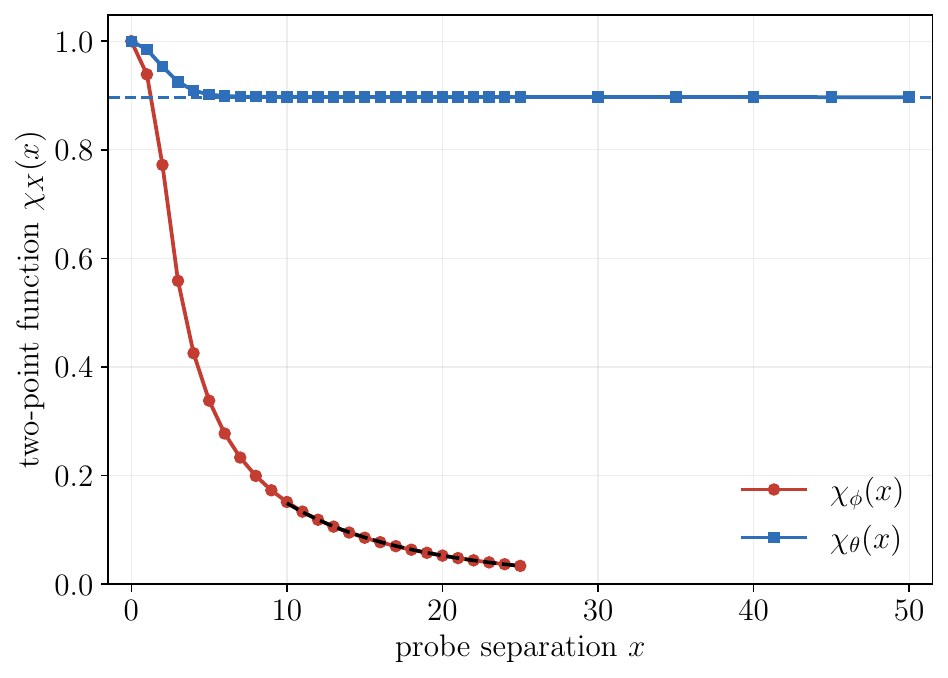}
    \caption{Distance dependence of the representative unit-charge correlation functions $\chi_{\theta,1}(x)=\langle e^{i\theta(x)}e^{-i\theta(0)}\rangle$ and $\chi_{\phi,1}(x)=\langle e^{i\phi(x)}e^{-i\phi(0)}\rangle$ at the QCP, obtained from the exact numerical solution of the saddle-point equations~\eqref{app:eq:source-saddle-phi} and~\eqref{app:eq:source-saddle-theta} on a periodic lattice. The plot shows that the unit-charge $\theta$ vertex exhibits long-range order, with $\ln\chi_{\theta,1}(x\to\infty)=A_{\theta,1}$, where $A_{\theta,1}\neq 0$ (blue dashed line), whereas $\chi_{\phi,1}(x)$ displays a nontrivial decay, $\log\chi_{\phi,1}(x)\propto -x^{2/3}$ (black dashed line). The results shown correspond to $\beta=1$ and the analytically continued limit $N=1$; the analytical results above give the charge dependence for general allowed $\beta$.} 
\label{fig:fig1}
\end{figure}

We solved Eqs.~\eqref{app:eq:source-saddle-phi} directly on a periodic $(\tau',x')$ lattice using Fourier spectral derivatives, with $\Delta x'=1$, $\Delta\tau'=2$, $\lambda_{4}=1$, and Gaussian packet width $\sigma=1$. We used a $150\times150$ lattice with periodic boundary conditions, corresponding to $L_{x'}=150$ and $L_{\tau'}=300$. The larger extent in the imaginary-time direction was chosen to account for the expected anisotropic scaling, Eq.~\eqref{app:eq:xi_x_xi_tau}. One complex variable $r_i$ was assigned to each spacetime site. At each Newton iteration, we reconstructed the full source-dependent Green's matrix, explicitly projected out the uniform critical zero mode, and evaluated the exact analytic Jacobian of the discretized saddle-point equations. Newton steps were stabilized using a backtracking line search with step sizes $\alpha=2^{-m}$, $m=0,\ldots,11$. For Eq.~\eqref{app:eq:source-saddle-phi}, the complex saddle was initialized with the variational bridge profile and phase $2\pi/3$. Solutions were accepted when the normalized residual fell below $5\times10^{-4}$. The reported actions and correlation functions were evaluated on the full nonuniform saddle without invoking a local or gradient expansion.

For the representative $\beta=1$ correlator $\chi_{\theta,1}$, we first solved the full source-deformed saddle-point equation~\eqref{app:eq:source-saddle-theta} and verified that, for separations up to $x\simeq 10$, the action evaluated on the exact nonuniform saddle agrees, within numerical accuracy, with that obtained from the bare saddle, $r=0$; see Fig.~\ref{fig:theta-nonlinear-saddle}. We used the same numerical setup as for the $\chi_{\phi,1}$ calculation: $\Delta x'=1$, $\Delta\tau'=2$, $\lambda_{4}=1$, and Gaussian packet width $\sigma=1$. The calculations were performed on a periodic $150\times150$ lattice with $L_{x'}=150$ and $L_{\tau'}=300$. The normalized residual tolerance was chosen to be $10^{-7}$. At larger separations, the $z=2$ scaling requires the imaginary-time extent to grow parametrically faster than the spatial extent, $L_{\tau'}\sim L_{x'}^{\,2}$, in order to resolve the infrared scale $\omega\sim k^{2}$. A full solution of the saddle equation on such lattices, which requires repeatedly constructing and inverting the source-dependent Green's matrix, becomes computationally prohibitive. We therefore evaluated the large-distance $\chi_{\theta,1}$ action at the bare saddle $r=0$, whose agreement with the exact saddle had already been established at accessible separations. For the data shown in the Figure~\ref{fig:fig1}, we used
\[ 
    L_{x'}=150,\qquad L_{\tau'}=L_{x'}^{\,2}=22500,\qquad \Delta x'=1,\qquad \Delta\tau'=2.
\]
At $r=0$, the action was evaluated directly in Fourier space using separable spatial and temporal sums, without constructing the full Green's matrix. This anisotropically scaled lattice resolves the long-wavelength $z=2$ modes and reveals the large-distance saturation of $\chi_{\theta,1}(x)$.

\section{Bosonization of the lattice model}
\label{app:bosonization_appendix}
After linearizing the dispersion around $\pm k_F$, one can split the lattice fermions into left- and right-chiral components: $c_{i\sigma} = \sqrt{a}\left(e^{ik_Fx}\psi_{+\sigma}(x) + e^{-ik_Fx}\psi_{-\sigma}(x)\right)$. These can then be written using emergent bosonic fields using the bosonization dictionary as 
\begin{align}
    \psi_{\pm\sigma} = \frac{\eta_{\pm\sigma}}{\sqrt{2\pi\alpha}}e^{\pm i\sqrt{4\pi}\phi_{\pm \sigma}}\,, \quad {\eta_{\pm \sigma}^{\dagger} = \eta_{\pm \sigma} \,, \quad \{\eta_{s\sigma}, \eta_{s'\sigma'}\} = 2 \delta_{ss'} \delta_{\sigma\sigma'}\,.}
\end{align}
The chiral components of the bosonic fields satisfy the commutation relations
\begin{equation}
    {[\phi_{s\sigma}(x), \phi_{s'\sigma'}(x')] = \frac{is}{4} \delta_{ss'} \delta_{\sigma\sigma'} \mathrm{sgn}(x-x') \,.}
\end{equation}
For each spin flavour, the chiral components $\psi_{\pm}$ can be recombined into an effective field and its dual : $\phi = \phi_++\phi_-$ and $\theta = \phi_+-\phi_-$. A convenient choice of basis transformation is $\phi_{c/s} = \frac{1}{\sqrt{2}}(\phi_\uparrow \pm \phi_\downarrow)$, and likewise for the dual fields, resulting in the Hadamard transform relation: 
\begin{align}
    \frac{1}{\sqrt{2}}
    \begin{bmatrix}
        \phi_c \\ \phi_s \\ \theta_c \\ \theta_s
    \end{bmatrix} = 
    \frac{1}{2}
    \begin{bmatrix}
        1 & 1 & 1 & 1 \\ 1 & 1 & -1 & -1 \\
        1 & -1 & 1 & -1 \\ 1 &-1 & -1 & 1
    \end{bmatrix}
    \begin{bmatrix}
        \phi_{+\uparrow}\\ \phi_{-\uparrow} \\ \phi_{+\downarrow} \\ \phi_{-\downarrow}
    \end{bmatrix} \,.
\end{align}
The bosonization map can be applied to lattice quantities as 
\begin{align}
    \rho(x) &= {\rho_0 + } \sqrt{\frac{2}{\pi}}\partial_x\phi_c + C\,e^{i2k_Fx}\mathcal{O}_{\mathrm{CDW}} + \mathrm{h.c.} \nonumber \\
    S^z(x) &= \frac{1}{\sqrt{2\pi}}\partial_x\phi_s + C\,e^{2ik_Fx}\mathcal{O}_{\mathrm{SDW}} + \mathrm{h.c.} 
\end{align}
These relations can be used to determine the Luttinger parameters $K_c$ and $K_s$ respectively, by the slope of the Fourier transform near $q=0$. Using the relation that 
\[
    \int_{-\infty}^\infty dx \frac{e^{-iqx}}{x^2} = -\pi |q| \,, 
\]
one obtains that in the LL phase, the slope of $\mathcal{F}\avg{\rho(x)\rho(0)}_c$ and $\mathcal{F}\avg{S^z(x)S^z(0)}_c$ at $q=0$ give $\frac{K_c}{\pi}$ and $\frac{K_s}{4\pi}$ respectively. 
Additional bosonized operators are given in Tab.~\ref{tab:bosonized_qtys}, and the Luttinger liquid character of the paramagnetic phase is verified in Fig.~\ref{fig:PMisLL}.
\begin{table}[h!]
    \centering
    \begin{tabular}{c|c|c|c|c|c}
         Quantity $\mathcal{O}$ & Lattice operator &Uniform comp. & $k_F$ & $2k_F$ & $\langle \mathcal{O}(x)\mathcal{O}^\dagger(0)\rangle$ in LL\\
         \hline\hline
         charge & $\sum_{\sigma}c^\dagger_{i\sigma}c^{\phantom\dagger}_{i\sigma}$ & $\sqrt{\frac{2}{\pi}}\partial_x\phi_c$ & - & $e^{i\sqrt{2\pi}\phi_c}\cos{\sqrt{2\pi}\phi_s}$ & $-\frac{K_c}{\pi^2x^2} + C|x|^{-(K_c+K_s)}\cos{2k_Fx}$\\
         \hline
         Spin-z & $\frac{1}{2}\sum_{\sigma}c^\dagger_{i\sigma}\sigma^zc^{\phantom\dagger}_{i\sigma}$ & $\sqrt{\frac{1}{2\pi}}\partial_x\phi_s$ & - & $e^{i\sqrt{2\pi}\phi_c}\sin{\sqrt{2\pi}\phi_s}$ & $-\frac{K_s}{4\pi^2x^2} + C|x|^{-(K_c+K_s)}\cos{2k_Fx}$\\    
         \hline
         $S^+$ & $c^\dagger_{i\uparrow}c^{\phantom\dagger}_{i\downarrow}$ & $e^{-i\sqrt{2\pi}\theta_s}\cos\sqrt{2\pi}\phi_s$ & - & $e^{-i\sqrt{2\pi}\theta_s}e^{i\sqrt{2\pi}\phi_c}$ & 
         $A|x|^{-(K_s + 1/K_s)} + B|x|^{-(K_c + 1/K_s)}\cos{2k_Fx}$ \\
         \hline
         electron & $c_{\uparrow}$ & - & $e^{i\sqrt{\frac{\pi}{2}}(\phi_c+\phi_s+\theta_c+\theta_s)}$& - & $C|x|^{-\frac{1}{4}(K_c + K_s + 1/K_c + 1/K_s)} \cos{k_F x}$ 
    \end{tabular}
    \caption{Bosonized representation of some common lattice quantities.}
    \label{tab:bosonized_qtys}
\end{table}
\begin{figure}[t!]
    \centering
    \includegraphics[width=0.9\linewidth]{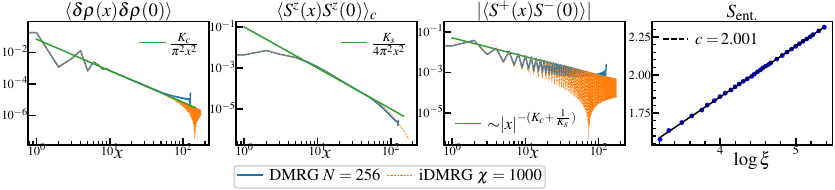}
    \caption{The paramagnetic phase can be well approximated by a Luttinger liquid, as seen by an agreement in both finite dmrg, infinite dmrg and bosonization. The data is for $\Delta V = 3.2$. Shown also is the central charge, obtained for $\Delta V = 3.0$, but is qualitatively the same throughout the phase.}
    \label{fig:PMisLL}
\end{figure}

In a spin-only model, such as in Ref.~\cite{Nahum2025}, the $S^+$ operator on the lattice maps directly to the vertex operator of the dual field in the bosonized description, but as can be seen from Tab.~\ref{tab:bosonized_qtys}, both its uniform and $2k_F$ components pick up additional contributions from other vertex operators of $\phi_{s/c}$ fields that cause it to decay at long distances, even at the critical point, where the slowest decay is $\sim|x|^{-K_c}\cos{2 k_F x}$.  
We observe instead that the spin-2 operator $S^+_iS^+_{i+1}$ on the lattice is the operator that depends only on the $\theta_s$ vertex in the effective description. This operator maps to
\begin{align*}
    \OO_2 &= {c^\dagger_{\uparrow}c_\downarrow(x) c^\dagger_\uparrow c_\downarrow(x+a)} \nonumber \\
    &= a^2\left[\psi^\dagger_{+\uparrow}\psi^{\phantom\dagger}_{+\downarrow} + \psi^\dagger_{-\uparrow}\psi^{\phantom\dagger}_{-\downarrow} + e^{2ik_Fx}\psi^\dagger_{-\uparrow}\psi^{\phantom\dagger}_{+\downarrow} + e^{-2ik_Fx}\psi^\dagger_{+\uparrow}\psi^{\phantom\dagger}_{-\downarrow} \right](x) \nonumber \\
    &\quad\quad\quad\left[\psi^\dagger_{+\uparrow}\psi^{\phantom\dagger}_{+\downarrow}+ \psi^\dagger_{-\uparrow}\psi^{\phantom\dagger}_{-\downarrow} + e^{2ik_F(x+a)}\psi^\dagger_{-\uparrow}\psi^{\phantom\dagger}_{+\downarrow} + e^{-2ik_F(x+a)}\psi^\dagger_{+\uparrow}\psi^{\phantom\dagger}_{-\downarrow} \right](x+a)
\end{align*}
The terms that depend only on $\theta$ or $\phi$ alone will have net chirality zero. In the uniform component, these are 
\begin{align}
    T_1 &= \psi^\dagger_{+\uparrow}\psi^{\phantom\dagger}_{+\downarrow}(x)\psi^\dagger_{-\uparrow}\psi^{\phantom\dagger}_{-\downarrow}(x+a) 
    + \psi^\dagger_{-\uparrow}\psi^{\phantom\dagger}_{-\downarrow}(x)\psi^\dagger_{+\uparrow}\psi^{\phantom\dagger}_{+\downarrow}(x+a) \quad, \text{and} \nonumber \\
    T_2 &= e^{-i2k_Fa}\psi^\dagger_{-\uparrow}\psi^{\phantom\dagger}_{+\downarrow}(x)\psi^\dagger_{+\uparrow}\psi^{\phantom\dagger}_{-\downarrow}(x+a)
    + e^{i2k_Fa} \psi^\dagger_{+\uparrow}\psi^{\phantom\dagger}_{-\downarrow}(x)\psi^\dagger_{-\uparrow}\psi^{\phantom\dagger}_{+\downarrow}(x+a) \,.
\end{align}
\noindent The terms in $T_1$ and $T_2$ have opposite Klein factor parities, and they can be summed to obtain 
\begin{align}
    \OO_2 = \eta \, \frac{4a^2\sin^2{k_Fa}}{4\pi^2\alpha^2}e^{-i\sqrt{8\pi}\theta_s} = \eta \, \frac{4a^2\sin^2{k_Fa}}{4\pi^2\alpha^2}e^{-2i\theta} \,, \quad {\eta = \eta_{+\uparrow} \eta_{+\downarrow} \eta_{-\uparrow} \eta_{-\downarrow} \,, \quad \eta^2 = 1\,,}
\end{align}
which, in the LL phase, decays asymptotically as $\sim|x|^{-4/K_s}$, and develops true long range order at the critical point. On the other hand, a pure $\phi_s$ vertex can be achieved with the local operator $S^+_iS^-_{i+1}\sim\cos{\sqrt{8\pi}\phi_s}={\cos(4\pi\phi)}$, corresponding to the conjugate vertex with $|\beta|=4\pi$.

\section{DMRG results}
In this section we show extended results for the DMRG numerics of the lattice model considered in the main text.

\begin{figure}[h!]
    \centering
    \includegraphics[width=0.9\linewidth]{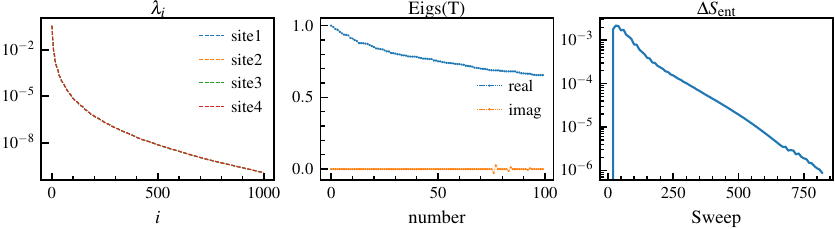}
    \caption{iDMRG convergence checks at the critical point: (a) Squared Schmidt coefficients, showing a rapid decay. (b) The first 100 eigenvalues of the transfer matrix. It can be noted that there is no exact degeneracy between the first two eigenvalues. (c) Entanglement entropy change as a function of each iDMRG sweep indicating a uniform convergence to the final obtained state.}
    \label{fig:dmrg_diagnostics}
\end{figure}

\subsubsection{DMRG parameters}
In this section, we specify the parameters of the DMRG for ease of reproducibility of the data presented in the main text. In all cases, particle number conservation was imposed, but not conservation of $S^z$.
Finite DMRG calculations were performed using the ITensors.jl library~\cite{ITensor,ITensor-r0.3}. The data presented in the paper was collected up to a bond dimension of $\chi=500$ for all system sizes and all couplings. Starting from the paramagnetic state, a scan of the quantity $\Delta V$ was performed by using the MPS at coupling $\Delta V$ as the starting state in the DMRG procedure for the coupling $\Delta V$ + $\delta \Delta V$, with $\delta \Delta V = 0.05$. Convergence was reached when successive DMRG iterations had an energy difference lower than a set threshold, chosen in the range $10^{-5} - 10^{-6}$. The variance of the energy in the ground state, $\mathrm{Var} E = \avg{\hat{H}^2} - \avg{\hat{H}}^2$, with $\hat{H}$ being the Hamiltonian, was monitored in the final obtained state. This quantity being close to zero is often used as a metric of how close the obtained state is to an eigenstate~\cite{mcculloch2008infinite}, and typically was $\sim10^{-4}$. Data was collected for higher bond dimension (up to $\chi=1200$) for the purpose of verifying the result, but did not change the numerical values of the correlation functions, or the variance of the ground state energy significantly.  

The iDMRG calculations were performed using the TenPy package~\cite{tenpy2024}. We pick a 4-site unit cell, which is repeated. Correlation functions are obtained using the eigenvalues of the transfer matrix. Even though the algorithm works in the infinite system size limit, the presence of a finite bond dimension imposes a finite correlation length $\xi$ in the system~\cite{pollmann2009theory}, $\xi/a$ is related to the bond dimension $\chi$ by a polynomial relation. All iDMRG correlation functions for physical quantities shown in the main text were calculated up to a distance of $0.9\xi$ for each corresponding bond dimension. 

As a check of convergence at the critical point, in Fig.~\eqref{fig:dmrg_diagnostics} we show the rapid decay of the square of the Schmidt coefficients, defined through the singular value decomposition across a bond as $\ket{\psi} = \sum_i\sqrt{\lambda_i}\ket{m_i}\otimes\ket{n_i}$. Defined this way, eigenvalues of the reduced density matrix are then $\lambda_i$.

\subsubsection{Nature of the phase transition}
The phase transition studied in the main text is of a second order nature. This is seen by the continuous vanishing of the net magnetization, and a continuous variation of the energy on tuning $\Delta V$ across the phase transition. see fig.~\eqref{fig:magacross_crit}.
\begin{figure}[h!]
    \centering
    \includegraphics[width=0.45\linewidth]{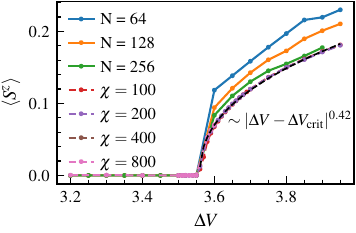}
    \caption{The total longitudinal magnetization density in the ferromagnetic phase smoothly goes to zero at the critical point, characteristic of a second order phase transition. The solid lines indicate finite size DMRG for increasing system sizes, and the dashed lines indicate iDMRG simulations at different bond dimensions. At large bond dimension, the correlation length is large enough that the iDMRG does not converge in the ferromagnetic phase, owing to the presence of phase separation. The black dashed line shows a three-parameter fit to the magnetization for the iDMRG data, taken at $\chi=200$.}
    \label{fig:magacross_crit}
\end{figure}
\subsubsection{Phase separation in the ferromagnetic state}
On crossing the critical point into the immediate ferromagnetic state, we see phase separation through a modulation of the overall density throughout the size of the finite size system, see fig.~\eqref{fig:supl_phase_sep}. Phase separation on the ordered side of the transition has also be predicted by Nahum in Ref.~\cite{Nahum2025}. However, very deep into the ferromagnetic phase at very large $\Delta V$ when the magnetization saturates, we observe once again a Luttinger liquid phase with a gaped spin mode, i.e. central charge 1.
\begin{figure}[h!]
    \centering
    \includegraphics[width=0.45\linewidth]{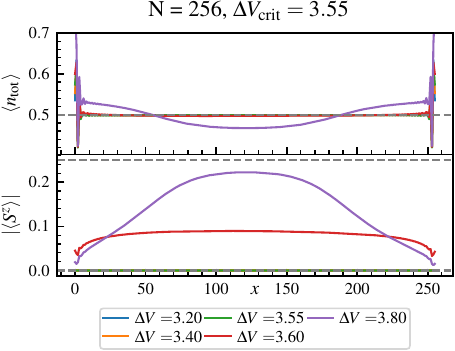}
    \caption{While the LL states at the critical point have complete translational invariance, in the ferromagnetically ordered state, the phase separation phenomenon is observed. For large enough system size, here $N=256$, the system rearranges its density into regions with a high and low magnetization respectively. }
    \label{fig:supl_phase_sep}
\end{figure}

\subsubsection{Details of the fitting procedure and finite entanglement scaling for $C(x)$}
\begin{figure}[h!]
    \centering
    \includegraphics[width=0.7\linewidth]{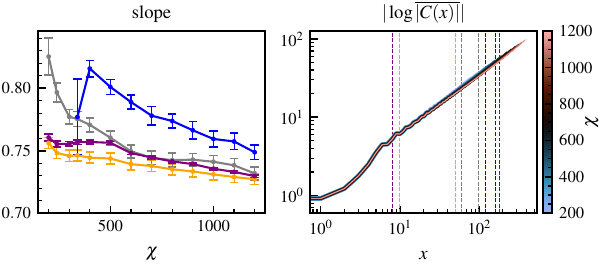}
    \caption{Finite entanglement scaling analysis at the critical point. The slope of the fit of $\log|\log\overline{|C(x)|}$ vs $\log x$ moves in the direction of the large-$N$ prediction of $2/3$ with increasing bond-dimension. The slope in the fit window of each color is shown as a function of bond-dimension. Each of the small windows are of length $40$ sites. Error bars denote the standard deviation of the data points from the fit line.}
    \label{fig:finite_ent_scaling}
\end{figure}
While an iMPS works with infinite boundary conditions and is useful to eliminate boundary effects completely, there is still a finite correlation length introduced due to having a finite bond dimension. Hence, correlation functions decay exponentially at length scales sufficiently greater than the correlation length. In other words, an iMPS at a finite bond-dimension behaves like a finite MPS of size comparable to the correlation length, but at an infinite bond-dimension. These corrections can fortunately be mitigated by finite entanglement scaling~\cite{pollmann2009theory}.

There are additional sources of fitting error in Fig.~\eqref{fig:finite_ent_scaling}. In order to extract the exponent $\alpha$ in the functional form $y=e^{-Cx^\alpha}$, one has to fit $\log(-\log y)$ vs $\log x$. This converts a decaying function into a rapidly growing one, thereby amplifying small numerical errors. 
{Furthermore, there are logarithmic corrections to the stretched exponential decay, with the true functional form being $\mathcal{N}|x|^{-\beta}\exp{(-C|x|^\alpha)}$.}
We find that averaging over a $4$-site window substantially smooths the data and suppresses the $k_F$ oscillations. The variance can be further reduced by increasing the fitting window, provided that it remains well below the correlation length. The results are shown in Fig.~\ref{fig:finite_ent_scaling}. At relatively short distances, while still well above the lattice scale (orange window), the fitted slope exhibits the weakest dependence on bond dimension. At longer distances (blue window), the slope shows a stronger negative drift with increasing bond dimension, consistent with the expected finite-entanglement effects.

\subsubsection{Spin-Spin correlation function at the critical point}
\begin{figure}[h!]
    \centering
    \includegraphics[width=0.45\linewidth]{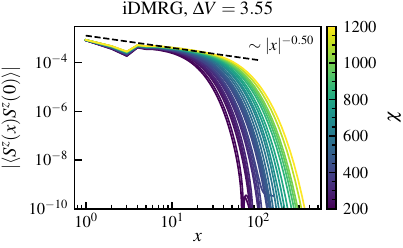}
    \caption{Correlation function for $S^z$ operator at the critical point from iDMRG, showing a power law scaling, strongly deviating from the Luttinger liquid $z=1$ result of $x^{-2}$. The data shown superposes the calculated correlation function at increasing bond dimension.}
    \label{fig:SzSz_at_crit_idmrg}
\end{figure}
Shown in Fig.~\eqref{fig:SzSz_at_crit_idmrg} is the two-point correlation function of the $S^z$ operator at the critical point. This is the lattice operator that probes correlation functions of $\partial_x\phi$ in the continuum field theory. In the iDMRG numerics, this operator has a very short correlation length in its sector at the critical point. This is seen in Fig.~\eqref{fig:SzSz_at_crit_idmrg} in the form of a crossover to exponential decay, whose onset gets pushed to larger distance with increasing bond dimension.
In the Luttinger liquid phase, $\langle S^z(x)S^z(0)\rangle$ should decay as $|x|^{-2}$ at long distances, as shown in Fig.~\eqref{fig:PMisLL}, and Tab.~\eqref{tab:bosonized_qtys}, consistent with $z=1$, as discussed in the main text. At the critical point, it picks up an anomalous exponent contribution. It must be cautioned that the fit value $0.5$ as seen in Fig.~\eqref{fig:SzSz_at_crit_idmrg} is not sufficient for accurately calculating the anomalous exponent with high precision, as the power-law doesn't extend for more than a decade, even at the largest calculated bond dimension of $\chi=1200$.
\end{widetext}

\end{document}